\documentclass{article}
\usepackage{microtype}
\usepackage{graphicx}
\usepackage{booktabs}
\usepackage[dvipsnames]{xcolor}
\usepackage{hyperref}
\usepackage[accepted]{mlsys2025}
\usepackage{etoolbox}
\makeatletter
\renewcommand{\Notice@String}{}
\patchcmd{\printAffiliationsAndNotice}{\ \\ \Notice@String}{}{}{\PackageWarning{header}{affiliation notice patch failed}}
\hypersetup{pdfsubject={}}
\AtBeginDocument{\chead{\small\bf\@mlsystitlerunning}}
\makeatother
\usepackage{multirow}
\usepackage{tikz}
\usepackage{amsmath}
\usepackage{amssymb}
\usepackage{siunitx}
\usepackage[abbreviations]{foreign}
\usepackage{xspace}
\usepackage[inline]{enumitem}
\usepackage[algo2e,ruled,lined,boxed,commentsnumbered, noend, linesnumbered, procnumbered]{algorithm2e}
\usepackage{cleveref}

\renewcommand{\cite}{\citep}

\newcommand{\doi}[1]{doi: \href{https://doi.org/#1}{\nolinkurl{#1}}}

\providecommand{\Description}[1]{}

\usepackage{framed}
\definecolor{highlightcolor}{RGB}{232, 244, 232}
\newenvironment{highlight}{\colorlet{shadecolor}{highlightcolor}\begin{shaded}\noindent}{\end{shaded}}

\usepackage{pifont}%

\usepackage{ifthen}

\ExplSyntaxOn
\seq_new:N \l__zzz_words_seq
\cs_new_protected:Npn \zzzhighlight #1
  {
    \seq_set_split:Nnn \l__zzz_words_seq { ~ } {#1}
    \seq_map_indexed_inline:Nn \l__zzz_words_seq
      {
        \int_compare:nNnF {##1} = { 1 }
          { { \color{yellow} \leaders \hrule height \ht\strutbox depth \dp\strutbox
              \hskip \fontdimen2\font plus .5em minus \fontdimen4\font } }
        { \setlength{\fboxsep}{0pt} \colorbox{yellow}{\strut ##2\/} }
      }
  }
\definecolor{jchengreen}{RGB}{0,110,60}

\ExplSyntaxOff
\newboolean{showcomments}
\setboolean{showcomments}{false}
\ifthenelse{\boolean{showcomments}}
{ \newcommand{\mynote}[3]{
		\fbox{\bfseries\sffamily\scriptsize#1}
		{\small$\blacktriangleright$\textsf{\emph{\color{#3}{#2}}}$\blacktriangleleft$}}
	\newcommand{\zzz}[1]{{\sffamily\itshape\zzzhighlight{#1}}\xspace}}
{ \newcommand{\mynote}[3]{}
	\newcommand{\zzz}[1]{}}

\ifthenelse{\boolean{showcomments}}
{ \newcommand{\OPEN}[1]{\par\medskip\noindent{\setlength{\fboxsep}{3pt}\fcolorbox{red}{red!15}{\parbox{0.94\columnwidth}{\textsf{\textbf{\color{red!60!black}{OPEN ISSUE:}} \emph{#1}}}}}\medskip}}
{ \newcommand{\OPEN}[1]{}}

\ifthenelse{\boolean{showcomments}}
{ \newcommand{\LATER}[1]{\par\medskip\noindent{\setlength{\fboxsep}{3pt}\fcolorbox{blue!50!black}{blue!8}{\parbox{0.94\columnwidth}{\textsf{\textbf{\color{blue!40!black}{LATER:}} \emph{#1}}}}}\medskip}}
{ \newcommand{\LATER}[1]{}}

\usepackage{acronym}

\makeatletter
\renewcommand*\AC@verridelabel[1]{}
\makeatother

\acrodef{MoE}{Mixture-of-Experts}
\acrodef{LLM}{large language model}
\acrodef{PEFT}{parameter-efficient fine-tuning}
\acrodef{EP}{expert parallelism}
\acrodef{TP}{tensor parallelism}
\acrodef{SLO}{service-level objective}
\acrodef{TTFT}{time-to-first-token}
\acrodef{TPOT}{time-per-output-token}
\acrodef{GEMM}{general matrix multiplication}
\acrodef{KV}{key-value}
\acrodef{HBM}{high-bandwidth memory}
\acrodef{EPLB}{expert-parallel load balancing}
\acrodef{LPLB}{linear-programming-based load balancer}
\acrodef{SM}{streaming multiprocessor}
\acrodef{LoRA}{low-rank adaptation}

\newcommand{\sysname}{\textsc{BubbleTea}}
\newcommand{\sysshortname}{BT}
\newcommand{\sysshort}{\sysshortname}
\newcommand{\sys}{\sysname\xspace}
\AddToHook{env/tikzpicture/begin}{\renewcommand{\sys}{\textbf{\sysname}\xspace}\renewcommand{\sysshort}{\textbf{\sysshortname}}}
\AddToHook{env/tabular/begin}{\renewcommand{\sys}{\textbf{\sysname}\xspace}\renewcommand{\sysshort}{\textbf{\sysshortname}}}
\newcommand{\llmstation}{\textsc{LLMStation}\xspace}
\newcommand{\harli}{\textsc{Harli}\xspace}

\newcommand{\waterfill}{\textsc{Waterfill}\xspace}
\newcommand{\uncoord}{uncoordinated colocation\xspace}
\newcommand{\Uncoord}{Uncoordinated colocation\xspace}
\newcommand{\cfgColo}{decode colocation\xspace}

\newcommand{\deepseekmoe}{\textsc{DeepSeek-V2-Lite}\xspace}
\newcommand{\sglang}{\textsc{SGLang}\xspace}
\newcommand{\flexllm}{\textsc{FlexLLM}\xspace}
\newcommand{\deltaserve}{\textsc{DeltaServe}\xspace}
\newcommand{\vllm}{\textsc{vLLM}\xspace}
\newcommand{\alpaca}{\textsc{Alpaca}\xspace}
\newcommand{\cfgOptColo}{\sys \emph{(without leveraging bubbles)}\xspace}
\newcommand{\cfgUnthr}{Unthrottled bubbles\xspace}
\newcommand{\cfgThrottled}{Throttled bubbles\xspace}

\graphicspath{ {figures/} }
\usepackage{tikz}
\usepackage[eulergreek]{sansmath}
\usepackage{pgfplots}
\usepackage{pgfplotstable}
\usepackage{cleveref}
\usepackage{comment}
\crefname{assumption}{assumption}{assumptions}

\pgfplotsset{compat=newest}
\usepgfplotslibrary{external,units,colorbrewer,groupplots,fillbetween,statistics}
\usetikzlibrary{patterns,shapes.misc}
\usetikzlibrary{positioning,arrows.meta,fit,backgrounds,calc,decorations.pathreplacing}

\newif\ifbuildplots
\buildplotsfalse
\ifbuildplots
	\tikzsetexternalprefix{figures/}
	\tikzset{external/mode=list and make}
	\newcommand{\inputplot}[1]{%
		{\sffamily\scriptsize
		 \tikzsetnextfilename{#1}\def\currentplot{#1}%
		 \input{plots/#1}\unskip}%
	}
\else
	\newcommand{\inputplot}[1]{\includegraphics{#1.pdf}}
\fi

\newcommand{\newgroupwidth}[2]%
{\expandafter\xdef\csname groupwidth#1\endcsname{#2}}

\def\currentplot{}
\newsavebox{\groupwidthbox}
\makeatletter
{\edef\groupnumber{#1}%
	\@ifundefined{groupwidth\currentplot}{\pgfmathsetlengthmacro{\mywidth}{\linewidth/\groupnumber}}%
	{\expandafter\let\expandafter\mywidth\csname groupwidth\currentplot\endcsname}%
	\begin{lrbox}{\groupwidthbox}%
		\tikzset{/pgfplots/width={\mywidth}}%
		\ignorespaces}%
	{\end{lrbox}%
	\usebox\groupwidthbox
	\@ifundefined{groupwidth\currentplot}%
	{\pgfmathsetlengthmacro{\mywidth}{\mywidth + (\linewidth - \wd\groupwidthbox)/\groupnumber}}{}%
	\tikzifexternalizing{}{%
		\immediate\write\@auxout{\string\newgroupwidth{\currentplot}{\mywidth}}%
	}}
\makeatother

\colorlet{bar1}{black!70}
\colorlet{bar2}{blue!70!black}
\colorlet{bar3}{green!55!black}
\colorlet{bar4}{red!70!black}
\colorlet{bar5}{violet!70}
\colorlet{bar6}{orange!85!black}

\definecolor{pastel1}{HTML}{8DA0CB}
\definecolor{pastel2}{HTML}{FC8D62}
\definecolor{pastel3}{HTML}{66C2A5}
\definecolor{pastel4}{HTML}{E78AC3}
\definecolor{pastel5}{HTML}{A6D854}
\definecolor{pastel6}{HTML}{FFD92F}
\definecolor{pastel7}{HTML}{E5C494}
\definecolor{pastel8}{HTML}{B3B3B3}

\colorlet{sysBubbleTea}{pastel2}
\colorlet{sysLLMStation}{pastel1}
\colorlet{sysNaive}{pastel3!70!black}
\colorlet{sysNaiveCap}{pastel3!45}
\colorlet{sysNaive40}{pastel3!75}

\colorlet{ladderA}{pastel8}
\colorlet{ladderB}{pastel7}
\colorlet{ladderC}{pastel6!50}
\colorlet{ladderD}{pastel4!65}
\colorlet{ladderE}{sysBubbleTea}

\newcommand{\plotheight}{1.55cm}

\pgfplotsset{compat=newest,
    scale only axis, width=\linewidth-40pt, height=\plotheight,
    every axis/.append style={
     semithick,
     ymajorgrids=true,
    },
    legend style={
     draw=none,
     fill=white,
     anchor=south,
     at={(0.5, 1)},
     font=\small,
    },
    cycle list/.define={lines}{
     {bar1},{bar2},{bar3},{bar4},{bar5},{bar6}},
    cycle list name=lines,
    cycle multiindex* list={
    	lines\nextlist
    	{dashed,solid,dotted,dashdotted,densely dashdotdotted}\nextlist
    	mark list*\nextlist
    },
    legend cell align=left,
    xtick align=outside,
    xtick pos=bottom,
    major tick length=1mm,
    /pgfplots/short line legend/.style={
     legend image code/.code={
         \draw[mark repeat=2,mark phase=2,##1]
             plot coordinates {
                 (0cm,0cm)
                     (0.2cm,0cm)
                     (0.4cm,0cm)
             };%
     }
    },
    /pgfplots/ybar legend/.style={
    /pgfplots/legend image code/.code={%
         \draw[##1,/tikz/.cd,yshift=-0.25em]
             (0cm,0cm) rectangle (4pt,6pt);},
    },
    /pgfplots/xbar legend/.style={
    /pgfplots/legend image code/.code={%
         \draw[##1,/tikz/.cd,yshift=-0.25em]
             (0cm,0cm) rectangle (4pt,6pt);},
    },
    cycle list/.define={bars}{
     {fill=bar1,draw=black},
     {fill=bar2,draw=black},
     {fill=bar3,draw=black},
     {fill=bar4,draw=black},
     {fill=bar5,draw=black},
     {fill=bar6,draw=black}},
    cycle list/.define={pastel bars}{
     {fill=pastel1,draw=black},
     {fill=pastel2,draw=black},
     {fill=pastel3,draw=black},
     {fill=pastel4,draw=black},
     {fill=pastel5,draw=black},
     {fill=pastel6,draw=black}},
    bar plot/.style={
     ybar,
     /pgf/bar width=8pt,
     cycle list name=bars,
     error bars/error bar style={black, thin},
     enlarge x limits=0.14,
     xtick=data,
     ymin=0,
    },
    unit code/.code={\si{#1}},
    tick label style = {font=\sansmath\sffamily},
    every axis label = {font=\sffamily},
    legend style = {font=\sffamily},
    label style = {font=\sffamily},
    select coords between index/.style 2 args={
    	x filter/.code={
    		\ifnum\coordindex<#1\def\pgfmathresult{}\fi
    		\ifnum\coordindex>#2\def\pgfmathresult{}\fi
    }},
	discard if not/.style 2 args={
		x filter/.code={
			\edef\tempa{\thisrow{#1}}
			\edef\tempb{#2}
			\ifx\tempa\tempb
			\else
			\def\pgfmathresult{inf}
			\fi
	}}
}

\usepackage{amsmath,amssymb,amsfonts,amsthm}
\usepackage{physics}
\usepackage{enumitem}
\usepackage{bbm}
\theoremstyle{definition}

\theoremstyle{remark}

\allowdisplaybreaks

\mlsystitlerunning{Harvesting Expert-Parallel Bubbles for Fine-Tuning}

\begin{document}

\twocolumn[
\mlsystitle{Harvesting Expert-Parallel Bubbles for Fine-Tuning}

\begin{mlsysauthorlist}
\mlsysauthor{Oana Balmau}{mcgill}
\mlsysauthor{Saurav Banka}{eth}
\mlsysauthor{Jiaxuan Chen}{mcgill}
\mlsysauthor{Anne-Marie Kermarrec}{epfl}
\mlsysauthor{Ramya Narasimha Prabhu}{epfl}
\mlsysauthor{Rafael Pires}{epfl}
\mlsysauthor{Mathis Randl}{epfl}
\mlsysauthor{Martijn de Vos}{epfl}
\end{mlsysauthorlist}
\mlsysaffiliation{mcgill}{McGill University}
\mlsysaffiliation{eth}{ETH Zurich}
\mlsysaffiliation{epfl}{EPFL}
\mlsyscorrespondingauthor{Ramya Narasimha Prabhu}{ramya.prabhu@epfl.ch}

\mlsyskeywords{Mixture-of-Experts, expert parallelism, LLM serving, GPU sharing, parameter-efficient fine-tuning, LoRA}

\vskip 0.3in

\acresetall
\begin{abstract}
  GPUs are expensive, yet latency-critical \ac{LLM} inference leaves part of them idle.
  \emph{Colocation} reclaims this capacity by running best-effort work, such as fine-tuning, alongside serving.
  Many leading \acp{LLM} now use a \ac{MoE} architecture, which adds a new source of idle time.
  Under expert parallelism, \ac{MoE} experts are spread across GPUs, and because the router sends uneven numbers of tokens to each GPU, lightly loaded GPUs wait for the most loaded one.
  Even after applying load balancing solutions, such \emph{bubbles} still take 8--13\% of prefill time.
  We present \sys, the first colocation system that leverages expert-parallel bubbles for fine-tuning.
  Unlike existing colocation systems, which make one scheduling decision for all GPUs per forward pass, \sys predicts each GPU’s bubble at every layer from the router’s token assignment, and fills it with short chunks of fine-tuning work that reuse the weights already loaded for serving.
  We also introduce compiler passes that automatically derive these chunks from the served \ac{LLM} as-is, without porting each model architecture to a separate training engine.
  Our evaluation on three \ac{MoE} models, served on four A100 GPUs, shows that fine-tuning throughput with \sys is 1.9--6.2$\times$ faster than with \textsc{LLMStation}, a state-of-the-art colocation system.
  \sys keeps p90 inference latency within 10\% of serving alone in 50 of 63 workloads, compared to 44 for \textsc{LLMStation}. On bursty production traces, \sys trains 3.3$\times$ faster than \textsc{LLMStation} and 14$\times$ faster than a stock trainer colocated under CUDA MPS, with p90 inference latency staying within 7\% of serving alone.

\end{abstract}

]

\printAffiliationsAndNotice{}

\section{Introduction}
\label{sec:intro}

\acresetall

GPU resources are a dominant cost of serving \acp{LLM}, motivating systems that improve accelerator utilization~\cite{orca,vllm,sarathi,distserve,splitwise,muxserve,aegaeon}. Latency-critical inference must meet stringent response-time \acp{SLO}, yet rarely uses every GPU resource at every instant.
Colocation systems reclaim this otherwise-idle capacity by running best-effort work, such as fine-tuning, alongside serving as long as its \acp{SLO} are preserved~\cite{orion,llmstation,harli}.

Many recent frontier \acp{LLM} use a \ac{MoE} architecture~\cite{deepseekv3,deepseekv4,qwen3,kimik2,gptoss,gemini15,gemini25}. An \ac{MoE} layer contains many feed-forward \emph{experts}, and a router sends each token to only a few of them~\cite{switch}, increasing model capacity without proportionally increasing per-token computation.
Since for large \acp{LLM} the total size of experts is beyond what a single GPU can store, such models are typically served with \ac{EP}, which places different experts on different GPUs~\cite{deepseekv3,vllm-ep-rfc}.

\ac{EP}, however, introduces a new source of underutilization.
Because token routing is input-dependent, GPUs receive different amounts of expert work.
Lightly loaded GPUs finish early and wait for the results of the most-loaded GPU before computation can proceed.
We call the time a GPU spends waiting for the most-loaded GPU a \emph{bubble}.
This load imbalance is well known and heavily optimized through techniques such as custom token routing objectives, strategic expert replication and placement, token redistribution~\cite{switch,gshard,expertchoice,deepseekv3,fastermoe,eplb,lplb,waterfill,harmoeny}, and specialized communication and expert kernels~\cite{deepep,pplxkernels,megablocks}.
We build on these mechanisms and ask the question they leave open:
\begin{highlight}
  \emph{Once load balancing and execution optimizations have done
  their job, how much idle GPU capacity remains, and can it be leveraged to perform useful work?}
\end{highlight}

\begin{figure}[t]
  \centering
  \inputplot{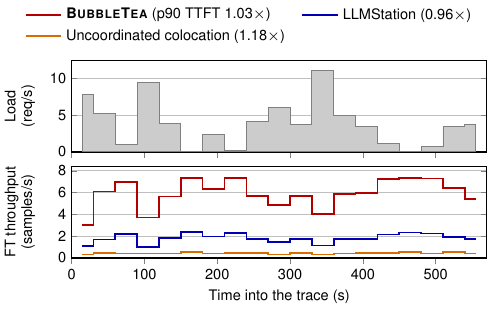}
  \caption{\textbf{\sys trains 3.3$\times$ faster than \llmstation and 14$\times$ faster than \uncoord on a production trace, while keeping p90 \ac{TTFT} within 3\% of serving alone.}
  Load (top) and fine-tuning throughput (bottom) of \sys and baselines over time on a \textsc{Qwen3-30B-A3B} model serving a production (Azure) trace at its original rate (with \SI{30}{\second} bins).
  Legend values give each system’s p90 \ac{TTFT} over the whole trace, relative to serving alone.}
  \label{fig:intro-trace}

\end{figure}

Our analysis in \S\ref{sec:characterization} answers the first half of this question and reveals that substantial idle capacity remains.
After applying state-of-the-art \ac{EP} load balancing techniques, bubbles still account for 8--13\% of prefill time.
The second half of the question, whether this capacity can perform useful work, is more challenging.
Existing colocation systems are natural candidates~\cite{flexllm,llmstation,harli,deltaserve}.
They place best-effort work, typically \ac{PEFT} such as \ac{LoRA}~\cite{lora}, on the same GPUs as inference.
However, they mostly exploit GPU underutilization during output generation (decode) and schedule work at the granularity of an inference or training step, whereas bubbles occur mostly during initial prompt processing (prefill), inside a single \ac{MoE} layer, and differ in length across GPUs.
Existing systems are therefore ineffective in leveraging such bubbles.

We introduce \sys, a colocation system that leverages bubbles to fine-tune \acp{LLM} with \ac{LoRA} while serving them.
The key insight behind \sys is that \emph{the length of a bubble is predictable}.
During inference, the router decides how many tokens each GPU receives before the experts run, and using this count, \sys predicts each GPU's bubble duration with a median error of \SI{8.9}{\micro\second}, about 7\% of the median bubble.
We introduce the first scheduler that admits fine-tuning work per GPU inside an \ac{MoE} layer, unlike existing colocation systems: \sys runs short chunks of fine-tuning work on each GPU only inside that GPU's predicted bubble window.

We also design a compiler that turns the served model into training code.
A serving engine only runs the model's \emph{forward pass}, which pushes tokens through the layers to predict the next one.
Training also needs a \emph{backward pass}, which propagates the error back through the model to compute the gradients that update the \ac{LoRA} adapter.
The compiler takes the serving engine's forward pass, derives the matching backward pass, and splits the resulting training step into chunks short enough to fit in a single bubble.
The generated code operates directly on the weights already loaded for serving, so \sys fine-tunes the model without keeping a second copy of its weights in GPU memory.
Existing colocation systems either train a second copy of the model~\cite{harli} or require porting each model architecture into their own training engine~\cite{flexllm,llmstation}, and they split training no finer than a layer or a serving iteration.
Our compiler instead works on the serving engine's model as-is, including its inference-only \ac{MoE} kernels and expert placement, so supporting a new architecture requires no hand-written kernels for backward passes.

We implement \sys and compare it against two baselines.
Since there are no colocation systems specifically designed for \ac{MoE}, we port an existing colocation system, \llmstation~\cite{llmstation}, to support inference with \ac{MoE} models, and also compare with \uncoord, a stock trainer running beside serving without coordination.
\Cref{fig:intro-trace} shows the fine-tuning throughput of \sys and both baselines while serving the \textsc{Qwen3-30B-A3B} \ac{LLM} on a bursty production trace.
Over the trace, \sys trains 3.3$\times$ faster than \llmstation and 14$\times$ faster than \uncoord, and it sustains this advantage throughout,
including during load bursts.
Meanwhile, the 90th-percentile \acf{TTFT} (p90 \ac{TTFT}) of \sys stays within 3\% of serving alone, whereas \uncoord increases it by 18\%.
Part of this gain comes from the fine-grained training runtime that bubbles require, which also makes better use of the decode and idle time that \llmstation already exploits.
Bubbles add the most at high load, when little decode and idle time is left for training, raising throughput by up to 27\% (\S\ref{sec:eval-ablation}).
Our further evaluation across three \ac{MoE} models and two production traces confirms that \sys turns idle time in \ac{MoE} serving into substantial fine-tuning throughput, while keeping both p90 \ac{TTFT} and \ac{TPOT} within 10\% of serving alone in 50 of 63 workloads.

This paper makes the following contributions:
\begin{itemize}
  \item \textbf{A characterization of expert-parallel bubbles}, showing that they survive
    load balancing, are predictable, and need per-GPU planning inside the \ac{MoE} layer (\S\ref{sec:characterization}).
  \item \sys, \textbf{a novel colocation system} that predicts each GPU’s bubble from the router’s decision, schedules short fine-tuning chunks only inside that window, and uses a compiler to derive those chunks automatically from the served \ac{LLM} (\S\ref{sec:design}).

  \item \textbf{An evaluation of \sys and baselines} on three \ac{MoE} models and two production traces that breaks the gains down by source of capacity and by mechanism (\S\ref{sec:evaluation}).
\end{itemize}

\section{Characterizing Expert-Parallel Bubbles}
\label{sec:characterization}

We next characterize expert-parallel bubbles and conclude that they are long enough, persistent enough, and predictable enough to schedule fine-tuning work inside.

\subsection{\ac{MoE} models and load imbalance}
An \ac{MoE} model replaces the dense feed-forward network in each transformer block with an \ac{MoE} layer, consisting of $E$ experts and a router that sends each token to its top-$k$ experts~\cite{switch,gshard}.
Large \acp{MoE} commonly combine \ac{TP} for dense computation with \acf{EP} for the experts.
\ac{EP} places different experts on different GPUs and routes individual tokens to the GPUs hosting the appropriate expert (\emph{dispatch})~\cite{deepseekv3,vllm-ep-rfc}.
In a \ac{TP}--\ac{EP} deployment, every GPU observes the same routing decision but executes only its local experts; the expert outputs are then combined across GPUs before the layer computation proceeds (\emph{gather}).

\begin{figure}[t]
  \centering
  \includegraphics[width=0.82\linewidth]{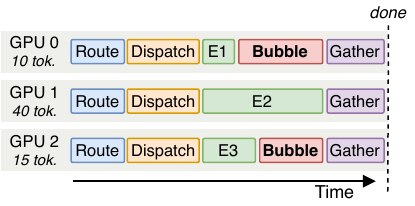}
  \caption{
    \textbf{Expert-parallel bubbles.}
    Token routing assigns different amounts of expert work
    to different GPUs. GPU~0 finishes its local expert computation (in green)
    before the more heavily loaded GPU~1 but cannot proceed until the
    cross-GPU dependency clears.
    We define this waiting interval (in red) as a \emph{bubble}.
  }
  \Description{
    Two expert-parallel GPUs execute different amounts of local expert
    work. GPU 0 finishes earlier than GPU 1 and waits until the
    cross-GPU dependency clears. This waiting interval is labeled the
    bubble.
  }
  \label{fig:anatomy}
\end{figure}

\Cref{fig:anatomy} shows a timeline of one \ac{MoE} layer executing on three GPUs.
Because routing is input-dependent, experts receive different numbers of tokens~\cite{switch,expertchoice,deepseekv3}.
Under \ac{EP}, this results in a per-GPU work imbalance: GPUs hosting more heavily loaded experts finish later, while lightly loaded GPUs finish earlier but cannot proceed as they wait for the processed tokens from other GPUs.
We call this GPU-local waiting interval a \emph{bubble}.
Specifically, a bubble is defined as the interval between completion of a GPU's local expert computation and the dependency that allows the computation to proceed.
We provide additional details on this in \S\ref{sec:appx-arch}.

\Ac{LLM} inference has two phases.
\emph{Prefill} processes the whole prompt and produces the first output token; its latency is measured by \ac{TTFT}, the time from a request's arrival to its first output token.
\emph{Decode} then generates the remaining tokens one at a time; its latency is measured by \ac{TPOT}, the time between consecutive output tokens.
Serving systems enforce \acp{SLO} on both.
In an \ac{MoE} model, the two phases give experts very different amounts of work.
During prefill, all prompt tokens pass through each \ac{MoE} layer in parallel, so expert computation is long and compute-bound.
During decode, each request contributes one token per step, so expert computation is short and memory-bandwidth-bound~\cite{orca,vllm}.
As we show next, this difference determines whether routing imbalance turns into bubbles long enough to leverage for useful work.

\subsection{The characteristics of bubbles}
\label{sec:char-obs}

To understand the characteristics of expert-parallel bubbles and how we can best leverage them, we run three experiments and instrument \ac{MoE} layer execution with CUDA events.
We measure, for each GPU and layer, the bubble duration.
For each experiment, we provide key details of the experiment setup, and provide additional details in \S\ref{sec:appx-setup-obs}.

\paragraph{Observation 1: Prefill creates bubbles we can leverage, and that grow
with prompt length.}
\label{sec:char-phase}
\label{sec:char-context}

Routing imbalance occurs during both prefill and decode, but whether a bubble can be leveraged depends on the resulting \emph{absolute} difference in expert computation time across GPUs.
We first measure routing skew, which we define as the most-loaded GPU's routed tokens divided by the mean.
On the \textsc{Qwen3-30B-A3B} model with experts split across two GPUs and no load balancing, and using \textsc{arXiv}-summarization prompts~\cite{cohan-etal-2018-discourse}, we find that this skew is 1.28$\times$ during prefill and 1.75$\times$ during decode.
Thus, token imbalance is more pronounced during decode than prefill on this dataset.
However, because decode expert computation is short, this larger imbalance yields only small bubbles.
Across batch sizes 1--128 and \textsc{arXiv}-summarization prompts of 32--512 tokens in length, while generating 30 tokens per request, the median bubble duration during decode is \SIrange{5}{25}{\micro\second} per layer.
This is insufficient opportunity to effectively schedule work (in \S\ref{sec:design} we use training chunks of \SI{50}{\micro\second}).
When performing prefill with \textsc{arXiv}-summarization prompts of length 8k, in contrast, the bubble length is \SI{0.29}{\milli\second} per layer, which is about 14\% of the layer's expert computation time.
\emph{Thus, prefill offers far more opportunities than decode to schedule work in bubbles.}

\begin{figure}[t]
  \centering
  \inputplot{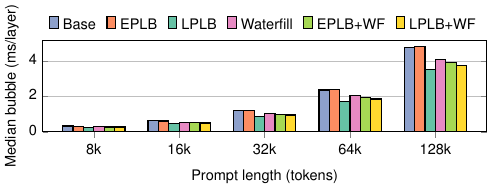}
  \caption{
    \textbf{Load balancing optimizations reduce bubbles but do not eliminate them.}
    The median per-layer bubble duration on \deepseekmoe, 8$\times$H100 GPUs, and with BF16 precision, for prompts with varying lengths, with and without (combinations of) load balancing optimizations.%
  }
  \label{fig:lb-sweep}
\end{figure}

The bubble duration during prefill also grows rapidly with the prompt length.
\Cref{fig:lb-sweep} shows the median bubble duration during prefill, for varying prompt lengths, on the \textsc{DeepSeek-V2-Lite} model, with and without load balancing optimizations.
In the \emph{Base} configuration (no optimizations), the median per-layer bubble duration increases substantially when increasing prompt length from 8k to 128k tokens.
The same trend also holds for other \acp{LLM} (\textsc{Qwen3-30B-A3B} and \textsc{Qwen1.5-MoE-A2.7B}, see \S\ref{sec:appx-context}).
\emph{Thus, longer prompts create more opportunities to fill bubbles.}

\paragraph{Observation 2: Load balancing optimizations do not eliminate bubbles.}
\label{sec:char-optimizations}

We next quantify how much bubble duration remains after applying the following three load balancing optimizations.
\begin{enumerate*}[label=(\roman*)]
    \item \Ac{EPLB} uses historical expert loads to place experts and replicate hot ones across GPUs~\cite{deepseekv3,eplb}, giving the runtime multiple expert replicas to which tokens can be sent.
    \item \ac{LPLB}~\cite{lplb} further balances routed tokens across those replicas at dispatch time using linear programming.
    \item \waterfill~\cite{waterfill} assigns shared-expert work to the least-loaded GPUs.
\end{enumerate*}
\waterfill composes naturally with the other two optimizations: \ac{EPLB} and \ac{LPLB} route tokens to physical expert replicas, then \waterfill balances the shared-expert workload.
We therefore also consider their combination.

We evaluate these mechanisms on the \deepseekmoe \ac{LLM} using the \sglang serving engine, 8$\times$H100 with experts split across eight GPUs, and with BF16 precision.
\Cref{fig:lb-sweep} shows the median bubble duration across prompt lengths.
Load balancing reduces the bubble duration substantially compared to \emph{Base} but does not eliminate it.
With prompt lengths of \SI{128}{k}, \emph{Base} exposes \SI{4.79}{\milli\second\per layer}, while \ac{LPLB}, the most competitive load balancing system, still leaves \SI{3.53}{\milli\second} of idle time per layer.
Load balancing is not free either: replicating experts consumes GPU memory, and redistributing tokens at dispatch time can increase communication volume.
For example, \waterfill raises median dispatch time from \SIrange{2.50}{2.65}{\milli\second\per layer} (\SI{6.0}{\percent}); correspondingly, \ac{LPLB}+\waterfill leaves a \SI{3.76}{\milli\second\per layer} bubble, compared to \SI{3.53}{\milli\second\per layer} with \ac{LPLB} alone.
\emph{Thus, minimizing token imbalance by itself does not necessarily eliminate bubbles} and these optimizations must
be weighed against their memory, communication, and runtime overhead.
Lower-precision execution does not close the gap either: our experiments in \S\ref{sec:appx-fp8} show that FP8 leaves the bubble essentially unchanged relative to BF16.

\paragraph{Observation 3: Serving-level colocation systems miss per-GPU bubbles.}
\label{sec:gap}
We next investigate whether existing inference--fine-tuning colocation systems can already exploit the bubbles occurring during prefill.
\llmstation~\cite{llmstation} and \harli~\cite{harli} primarily exploit decode and idle-time underutilization instead of prefill.
\flexllm~\cite{flexllm} and \deltaserve~\cite{deltaserve} can place training work alongside prefill, but only when the serving iteration itself leaves sufficient headroom.
As the prompt length increases, however, GPU compute saturates quickly~\cite{sarathi} and under a stringent serving \ac{SLO}, their planners see little or no capacity in which to admit training.

The limitation is not simply the granularity of their training work.
These systems measure slack once per serving iteration, using signals such as batch composition, iteration latency, predicted request latency, or aggregate engine headroom (see \Cref{tab:colocation_systems} in \S\ref{sec:appx-gap}).
That is reasonable when idle capacity is spread evenly across the GPUs.
\ac{EP}, however, breaks this: within an \ac{MoE} layer, the iteration as a whole looks saturated because it runs at the pace of its most heavily loaded GPU, while the other GPUs finish their experts early and sit idle waiting for it.
A planner with a single headroom budget per iteration cannot see this per-GPU idle time, so it concludes that prefill has no usable capacity even though substantial bubbles remain on individual GPUs. In a trace-driven experiment, a planner that knows the exact mean bubble of each layer but assigns the same budget to every GPU uses only about 1\% of bubble time, whereas planning per GPU from the current routing decision uses 43--45\% (Appendix~\ref{sec:appx-gap}).
Thus, \emph{leveraging expert-parallel bubbles requires a different admission signal}.
The scheduler must estimate slack separately for each GPU at each \ac{MoE} layer, using the current routing decision to determine which GPUs can safely execute work.

\DeclareRobustCommand{\dstep}[1]{{\tikzexternaldisable\tikz[baseline=(s.base)]
  \node[circle, draw, inner sep=0.5pt, font=\tiny\bfseries] (s) {#1};}}

\section{Design of \sys{}}
\label{sec:design}

Observations~1 and 2 establish that bubbles expose usable GPU capacity even after token load balancing. Observation~3 establishes two requirements for using this capacity:
\begin{enumerate*}[label=(\roman*)]
    \item admission must be decided per GPU and \ac{MoE} layer from the router state; and
    \item fine-tuning must expose work short enough to fit inside the resulting bubbles, which determines how much of each bubble can be used (Appendix~\ref{sec:appx-gap}).
\end{enumerate*}

\sys{} addresses these requirements in two phases.
During the \emph{offline phase}, a fine-tuning compiler transforms the \ac{LLM} already used by the serving engine into a fine-tuning runtime (\S\ref{sec:design-compiler}).
Specifically, it derives the \ac{LoRA} backward pass over the same base weights, makes cross-GPU dependencies explicit, and partitions the remaining computation into \emph{fine-tuning chunks}, which are independently admissible units of GPU work. Sequences of these chunks and their dependency waits are captured as \textsc{CUDA} graphs so that they can be efficiently launched.

\begin{figure}[t]
\centering
\resizebox{\columnwidth}{!}{\inputplot{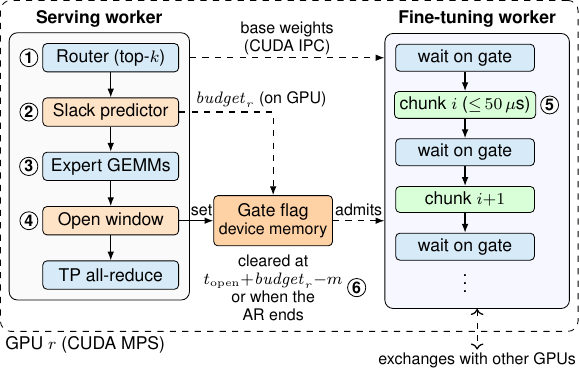}}
\caption{\textbf{The system architecture of \sys.} On each GPU, a serving worker and a fine-tuning worker run side by side and share the same copy of the base model weights.}
\label{fig:design-overview}
\end{figure}

During the \emph{online phase}, \sys uses each request's current \ac{MoE} router token assignment to predict the bubble duration on every GPU, then admits the precompiled fine-tuning work only inside those predicted bubbles (\S\ref{sec:design-window}).
As \Cref{fig:design-overview} shows, each GPU runs two processes: a \emph{serving worker} and a \emph{fine-tuning worker}.
A device-side gate lets the serving worker open and close an admission \emph{window} inside each GPU's bubble, while the fine-tuning worker keeps eligible work queued behind the gate.
The serving worker closes admission early enough that fine-tuning does not delay the serving-critical GPU, and the fine-tuning worker keeps work ready despite cross-GPU dependencies.

\begin{figure}[t]
\centering
\resizebox{\columnwidth}{!}{\inputplot{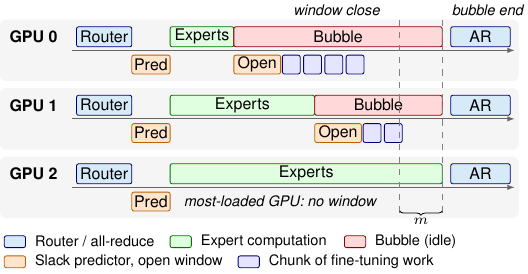}}
\caption{\textbf{Scheduling fine-tuning work with \sys on one \ac{MoE} prefill layer.}
Uneven expert work gives each GPU a different predicted bubble.
GPUs that finish early (GPU 0 and 1) execute queued chunks of work until shortly before the bubble ends; the most-loaded GPU (GPU 2) admits none.}
\label{fig:bubble_schedule}
\end{figure}

\subsection{\sys{} by example}
\label{sec:design-overview}

\Cref{fig:bubble_schedule} shows how \sys schedules fine-tuning work during one \ac{MoE} prefill layer across three GPUs.
\sys predicts each GPU's bubble directly from the router's expert assignment.
As soon as a GPU's experts finish, \sys runs queued chunks (derived with our compiler) back to back inside its bubble.
Because a running chunk cannot be preempted, \sys stops admitting new chunks a margin $m$ before the predicted end of the bubble, so that no chunk is still running when the all-reduce starts, which would prolong inference.

As a result, GPU~0 in \Cref{fig:bubble_schedule}, which has the largest bubble, executes more chunks than GPU~1, and GPU~2 has no bubble and runs no fine-tuning work at all.
Thus, fine-tuning progresses unevenly across GPUs.
In practice, we set $m$ to \SI{60}{\micro\second}, as described in Appendix \ref{sec:appx-margin}.

\subsection{Offline phase: deriving fine-tuning work}

\label{sec:design-compiler}

\paragraph{Why off-the-shelf autograd does not suffice.}
The technical challenge is to execute the \ac{LoRA} backward pass inside the serving deployment within bubbles.
Training frameworks such as \textsc{PyTorch} derive the backward pass automatically with \emph{autograd}: during the forward pass, they record each operation and keep the intermediate results it needs, and then traverse this record in reverse to compute the gradients.
Applying autograd to the served model fails in three ways.
\begin{enumerate*}[label=(\roman*)]
    \item \emph{The serving forward cannot be differentiated.} For speed, the serving engine (\vllm~\cite{vllm} in our case) runs expert computation, attention, and cross-GPU communication with specialized kernels that only produce outputs: they keep no intermediate results and have no backward pass, so autograd cannot run through them.
    Running autograd on a second, trainable copy of the model instead, as \textsc{torchtune} does in \uncoord, duplicates the \ac{LLM} weights.
    This takes another 15\,GB memory per GPU for \textsc{Qwen3-30B-A3B} that serving could otherwise use for the \ac{KV} cache.
    \item \emph{The backward pass is too coarse.} A bubble lasts under a millisecond and a kernel cannot be interrupted once started, so fine-tuning may only start work that finishes before the bubble ends.
    Autograd, however, runs the whole backward pass at once, one operator at a time, and a single expert matrix multiplication over a sample's tokens can already outlast a bubble. Even when work fits, larger units leave more of each bubble unused (\S\ref{sec:appx-gap}).
    \item \emph{The backward pass advances in lockstep.}
    Each GPU gets a different bubble, but the backward pass exchanges gradients across GPUs at every layer, and each exchange waits until all GPUs reach it.
    GPUs with large bubbles therefore sit idle at each exchange until the GPU with the least bubble time catches up, so most of the bubble capacity goes unused.
\end{enumerate*}

Existing colocation systems do not close this gap for expert-parallel \ac{MoE} models.
\llmstation splits training into suspendable tasklets, but at layer granularity with a host synchronization after each (\S\ref{sec:eval-e2e}).
Its fusion engine also assumes dense projections, which expert-parallel \ac{MoE} does not have, and fixed weight placement on GPUs, which \ac{EPLB} further breaks by moving experts across GPUs.

\paragraph{Deriving a bubble-compatible fine-tuning runtime.}
\sys{} addresses the three limitations above directly through the following three innovations.
\begin{enumerate*}[label=(\roman*)]
    \item \emph{Differentiate the serving \ac{LLM} in place.}
    It replaces inference-only operations with differentiable equivalents over the same serving weights and uses \textsc{AOTAutograd} to derive the backward graph, preserving the serving \ac{LLM}'s \ac{TP}/\ac{EP} layout and expert placement.
    \item \emph{Partition backward into bubble-sized work.}
    \sys makes cross-GPU exchanges explicit and partitions the remaining compute with a roofline cost model into chunks of at most \SI{50}{\micro\second}, splitting operators such as grouped expert \acp{GEMM} when necessary.
    Chunks are compiled with \textsc{Inductor} and sequences of gated chunks are captured in CUDA graphs to amortize launch overhead.
    Cross-GPU exchanges are annotated as communication chunks.
    \item \emph{Allow uneven progress across GPUs.}
    Instead of making every GPU wait for all others at each cross-GPU exchange, \sys lets a GPU wait only for the data it needs to keep progressing.
    One-sided mailbox messages make this possible: a GPU writes its data into a peer's mailbox in GPU memory and continues immediately.
    A GPU with larger bubbles can thus run ahead, while one with smaller bubbles falls a few chunks behind and catches up in later bubbles.
\end{enumerate*}
We provide additional details on our compiler passes in \S\ref{sec:appx-compiler}.

\subsection{Online phase: Predicting and admitting work}
\label{sec:design-window}

At runtime, \sys uses the current \ac{MoE} routing decision to predict how long the bubble on each GPU will last and then schedules the precompiled fine-tuning work only inside those bubbles.
We next explain this process.

\paragraph{Predicting each GPU's bubble.}
\label{sec:design-predict}
In an \ac{EP} deployment, every GPU observes the full top-$k$ assignment, so it can directly estimate the expert work assigned to every GPU.
\sys introduces a predictor that estimates bubble durations from this assignment (Step 2 in \Cref{fig:design-overview}).
The serving engine computes all experts on a GPU with one grouped matrix-multiplication kernel, which processes each expert's tokens in \emph{tiles}: fixed-size blocks of $M$ tokens that each take roughly the same time.
An expert $e$ with $n_e$ routed tokens therefore needs $\lceil n_e/M\rceil$ tiles, and the predictor sums these tile counts over the experts placed on each GPU.
The predicted bubble duration for GPU $g$ is then $\mathit{bubble}_g = \alpha(\max_j \mathit{tiles}_j - \mathit{tiles}_g)$, where $\alpha$ is the execution time of a single tile, so the most-loaded GPU has no bubble.
We derive $\alpha$ once per model and GPU configuration from a profiling run; porting the predictor to other hardware or \ac{GEMM} kernels only requires setting the corresponding tile height $M$ and re-estimating $\alpha$.

The predictor runs as a small GPU kernel immediately after routing and keeps its result in device memory for the admission gate.
Our predictor estimates bubble durations accurately and efficiently.
Its median error is \SI{8.9}{\micro\second}, about 7\% of the median bubble duration at a layer (\S\ref{sec:appx-predict}), and its serving-path overhead is at most 0.5\% (\S\ref{sec:eval-overheads}).
For prefills with little computation, however, the available bubbles are too small to justify the prediction and gating overhead.
For prompts with 2k tokens, enabling bubble filling reduces fine-tuning throughput by 8--11\% (\S\ref{sec:appx-threshold}).
\sys{} therefore disables bubble filling below 4k tokens and leaves those requests to conventional decode and idle-time colocation.

\paragraph{Admitting work inside the predicted bubble.}
Each GPU has a gate flag in device memory that the serving worker sets and clears.
The fine-tuning worker enqueues the precompiled CUDA graphs ahead of time; each fine-tuning chunk waits on this gate before it executes.
The wait is a device-side stream wait and consumes no GPU compute.
When a GPU finishes its local expert computation and reaches the layer's gather operation, the serving stream opens its gate (Step 4 in \Cref{fig:design-overview}), allowing already-queued fine-tuning chunks to execute in the bubble.
Opening the gate is a single flag write, so serving never waits for fine-tuning: fine-tuning chunks wait on serving, never the reverse, and run only while the GPU would otherwise sit idle in its bubble.

\paragraph{Keeping work ready.}
\label{sec:design-chunk}
A bubble is only useful if a chunk can start the moment it opens.
A chunk can start only when its gate is open and its inputs have arrived, and some inputs come from cross-GPU exchanges with a slower GPU.
If the host waited for each such message, the dependent chunk would reach the GPU only after a host round trip, which is a significant fraction of a sub-millisecond bubble.
\sys instead resolves these dependencies on the device (\emph{device-side readiness}).
The host queues all chunks ahead of time, and the stream waits on the sending GPU's mailbox flag before each dependent chunk.
This wait occupies no SMs, so the chunk starts as soon as both its input and its gate are ready, without host involvement.

We call the ordered sequence of compute and communication chunks for one fine-tuning sample a \emph{fine-tuning chain}.
A chain whose next chunk waits on a lagging GPU still cannot use an open bubble.
\sys therefore keeps a second, independent fine-tuning chain in flight (\emph{paired chains}), with its own stream, buffers, and mailboxes.
While one fine-tuning chain is blocked on an exchange, the other can run its chunks, and both chains' gradients are accumulated into one optimizer step.
Because two concurrent chains increase contention for memory bandwidth, \sys runs them concurrently only when prefill took less than 70\% of the last second.
Otherwise, it serializes their execution.

\paragraph{Closing the window early.}
\label{sec:design-optC}
Because an admitted chunk cannot be preempted, one admitted near the end of a bubble may delay the all-reduce and thus increase inference latency.
\sys therefore stops admitting new chunks a margin $m$ before the predicted end of each bubble (\emph{early close}).
If GPU $g$ opens its gate at time $t_{\mathrm{open}}$, it closes the gate at $t_{\mathrm{close}} = t_{\mathrm{open}} + \mathit{bubble}_g - m $, where $\mathit{bubble}_g$ is its predicted bubble length.
A side-stream kernel closes the gate at $t_{\mathrm{close}}$, while completion of the serving all-reduce also clears it as a safeguard if the true bubble ends earlier than predicted.
No window is opened for bubbles shorter than $m$, so the most-loaded GPU never admits work.
We use $m=\SI{60}{\micro\second}$, chosen to exceed a typical fine-tuning chunk and absorb prediction error (\S\ref{sec:appx-margin}).

The same precompiled fine-tuning work and device-side gate are reused for conventional decode and idle-time colocation.
Only the admission bound differs: during prefill, each GPU's gate is limited by its router-predicted intra-layer bubble; during decode steps and idle periods, serving leaves capacity unused, so the gate stays open for the whole step.
This is the iteration-level colocation that existing systems such as \llmstation~\cite{llmstation} already perform.

\section{Evaluation}
\label{sec:evaluation}

We evaluate \sys and baselines on three \ac{MoE} models and ask the following questions.
First, how much fine-tuning throughput does \sys obtain and what is the impact on inference latency, compared with an iteration-level colocation system (\llmstation) and with uncoordinated colocation (\S\ref{sec:eval-e2e})?
Second, where does the throughput in \sys come from, and what execution policies are needed to leverage
intra-layer idle time safely (\S\ref{sec:eval-ablation}, \S\ref{sec:eval-latency})?
Third, does \sys still provide  benefits once the serving engine balances expert load itself (\S\ref{sec:eval-eplb})?
Appendix~\ref{sec:appx-eval} details our experimental setup, and Appendix~\ref{sec:appx-experiments} reports more extensive experiments: all models and prompt lengths, fewer GPUs, a second production trace, the \uncoord \ac{SM}-cap sweep, the choice of design parameters, bubble utilization, and overheads. Appendix~\ref{sec:appx-gap} isolates the value of per-GPU planning with a trace-driven replay.

\subsection{Experimental setup}
\label{sec:eval-method}

\paragraph{Hardware and models.}
We run all experiments on a single compute node equipped with four \textsc{NVLink}-connected A100-SXM4-80GB GPUs, serving with \vllm in BF16 precision.
We evaluate three \ac{MoE} models: \textsc{Qwen3-30B-A3B}~\cite{qwen3}, which has only routed experts,
\textsc{Qwen1.5-MoE-A2.7B}~\cite{qwen1.5_moe} and \textsc{DeepSeek-V2-Lite}~\cite{deepseekv2}, which additionally have shared experts that process every token and therefore add the same work to every GPU.

\paragraph{Workloads and metrics.}
For inference, we use \textsc{arXiv} summarization prompts~\cite{cohan-etal-2018-discourse} with lengths of 1k--32k tokens and generate 32 output tokens for each prompt.
We set \vllm's maximum number of sequences to 256.
For this controlled workload, we determine three load regimes.
With a \emph{high} load, the arrival rate is just below the point at which serving alone starts queueing.
Medium and low load use lower rates, which we specify per experiment and workload in the appropriate plots.
We also replay two production traces, \emph{code} and \emph{conversation} from the Azure LLM inference trace~\cite{splitwise}, each over a 10-minute window with the trace's own arrival times.
For fine-tuning, we use the \alpaca dataset, consisting of \num{52000} instruction-demonstration prompts~\cite{alpaca}.

\paragraph{Baselines.}
We compare against \llmstation~\cite{llmstation}, a state-of-the-art system for colocating fine-tuning with \ac{LLM} inference on shared GPUs.
\llmstation transforms a fine-tuning job into a suspendable pipeline and interleaves it with inference at iteration granularity, batching tuning work alongside decode steps.
The system leverages the compute left unused by memory-bound decode iterations, and the idle periods in which no inference work is pending,  which makes it the strongest representative of serving-level colocation.
\flexllm~\cite{flexllm} and \deltaserve~\cite{deltaserve} instead place fine-tuning work alongside prefill, but only when their per-iteration planner sees headroom in the serving iteration. At our prompt lengths, prefill saturates GPU compute~\cite{sarathi}, and an iteration-level view cannot see the per-GPU slack inside \ac{MoE} layers (\S\ref{sec:gap}). Their planners therefore admit almost no work into prefill. Since their training progress is tied to these admissions, they also make little use of decode slack, which leaves them mostly the idle time that \llmstation already exploits.
This makes iteration-level colocation by \llmstation the more appropriate point of comparison.
We also use \emph{\Uncoord} as a baseline, which runs a stock \ac{LoRA} trainer beside \vllm.

\paragraph{Implementation.}
We implement \sys in Python in about 8k lines on \vllm~0.21.1 (v1 engine), \textsc{PyTorch}~2.11, and \textsc{Triton}~3.6.
The offline compiler builds on \texttt{AOTAutograd} and \texttt{Inductor}.
At runtime, hooks in \vllm's \ac{MoE} layer predict and open each window.
A separate trainer process per GPU runs the compiled chunks under CUDA MPS and accesses the serving weights through CUDA IPC, so they are not duplicated.

For \llmstation, we port its artifact to our vLLM/\textsc{PyTorch} versions while preserving its scheduling policy.
Because its released fusion engine assumes dense projections with fixed weight placement, which does not support expert-parallel \ac{MoE}, we use its separate \textsc{PyTorch} training path under CUDA MPS.
We evaluate several \ac{MoE} expert-execution implementations and report the fastest, using padded batched \acp{GEMM} through \textsc{cuBLAS}.
\llmstation's latency is normalized to the same serving engine with training disabled.
\emph{\Uncoord} runs a stock torchtune \ac{LoRA} trainer beside \vllm under
CUDA MPS, with its \ac{SM} share capped at 40\%.

\paragraph{Metrics.}
For fine-tuning, we report throughput in samples/s.
For serving, we report p90 \ac{TTFT} and \ac{TPOT}, normalized to serving alone, which capture prefill and per-token decode latency, respectively.
Each experiment is repeated three times and we average all results.

\begin{figure}[t]
  \centering
  \inputplot{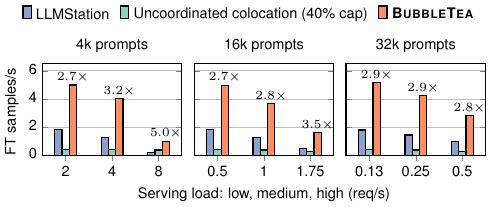}
  \caption{Fine-tuning throughput for \sys, \llmstation, and \uncoord (40\% cap) on \textsc{Qwen3-30B-A3B} across prompt lengths and serving loads. The labels above bars report \sys's fine-tuning throughput relative to \llmstation.}
  \label{fig:eval-e2e-tput}
\end{figure}

\subsection{End-to-end performance of \sys and baselines}
\label{sec:eval-e2e}

We measure end-to-end performance of \sys and baselines on the \textsc{Qwen3-30B-A3B} \ac{LLM}.

\paragraph{Steady-state workloads.}
\Cref{fig:eval-e2e-tput} shows fine-tuning throughput of \sys and baselines, for varying serving loads and 4k--32k prompt lengths.
Across all settings, \sys provides more fine-tuning throughput than \llmstation and trains 2.7--5.0$\times$ faster than \llmstation, with a median improvement of 2.9$\times$ (1.9--6.2$\times$ on all models, \S\ref{sec:appx-e2e}).
\Uncoord obtains a throughput of just 0.26--0.43 samples/s, 2.7--12.4$\times$ less than \sys, and lifting its \ac{SM} cap does not help: the off-the-shelf trainer is bound by its own per-expert kernel launches.
While \sys and \llmstation both fine-tune during decode and idle time, \sys uses fine-grained compiled chunks, which enables more efficient scheduling.

\begin{figure}[t]
  \centering
  \inputplot{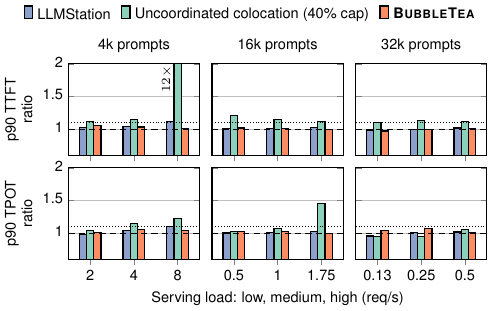}
  \caption{p90 \ac{TTFT} (top) and p90 \ac{TPOT} (bottom) ratio, relative to the corresponding serving-only execution, for \sys, \llmstation, and \uncoord (40\% cap) on \textsc{Qwen3-30B-A3B} across prompt lengths and serving loads. The dashed line marks serving alone and the dotted line a 10\% increase.}
  \label{fig:eval-e2e-lat}
\end{figure}

\Cref{fig:eval-e2e-lat} shows the p90 \ac{TTFT} and p90 \ac{TPOT} in the same settings as in \Cref{fig:eval-e2e-tput}, relative to a serving-only setting.
Across all 63 configurations, \sys keeps both p90 \ac{TTFT} and p90 \ac{TPOT} within 10\% of serving alone in 50 configurations (79\%), compared with 44 (70\%) for \llmstation.
The median increase of \sys is 4\% and 6\% for p90 \ac{TTFT} and p90 \ac{TPOT}, respectively.
\emph{\Uncoord} meets this 10\% bound in only 22 of the 63 configurations: it slows every serving step by a similar amount, which raises p90 latency by 10--20\% over serving alone when serving has slack, but at high load pushes serving past its saturation point, so requests queue and p90 \ac{TTFT} reaches 6--12$\times$ that of serving alone.
Thus, we conclude that \sys across most configurations achieves higher fine-tuning throughput without systematically increasing serving latency.

\paragraph{Production traces.}
We additionally replay two production traces with bursty, time-varying arrivals.
\sys shows 3.2--3.3$\times$ higher fine-tuning throughput than \llmstation while keeping p90 \ac{TTFT} and \ac{TPOT} within 7\% of serving alone.
\Cref{fig:intro-trace} in \S\ref{sec:intro} shows one representative trace.
Thus, even on production traces, \sys achieves superior fine-tuning throughput with low effect on inference latency.

\begin{figure}[t]
  \centering
  \inputplot{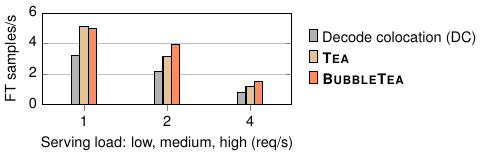}
  \caption{Fine-tuning throughput for \emph{decode colocation} (coarse-grained work scheduling), \textsc{Tea} (fine-grained work scheduling but not within bubbles), and \sys on \textsc{Qwen3-30B-A3B} with prompts of 8k length across serving loads.}
  \label{fig:eval-sources}
\end{figure}

\subsection{Where does \sys's throughput come from?}
\label{sec:eval-ablation}

Next, we investigate where the fine-tuning throughput in \sys comes from.
\Cref{fig:eval-sources} shows fine-tuning throughput for varying serving loads, and for different colocation mechanisms, while using prompts with length 8k.
We first compare \emph{\cfgColo}, which uses \llmstation-like scheduling at the granularity of an iteration, with a version of \sys, named \textsc{Tea}, that schedules work at finer granularity but does not schedule work in bubbles.
This comparison  isolates the benefit of \sys's fine-grained scheduling while holding the available serving capacity fixed to conventional decode and idle-time headroom.

\textsc{Tea} improves fine-tuning throughput by 36--60\% over \emph{\cfgColo}, with improvements more pronounced in the low load regime.
This highlights that \emph{the fine-grained scheduling mechanism in \sys is able to significantly increase fine-tuning throughput compared to conventional approaches.} \textsc{Tea} keeps p90 \ac{TTFT} and \ac{TPOT} within 10\% of serving alone in most configurations.

Next, we compare the fine-tuning throughput of \textsc{Tea} with that of \sys that additionally schedules fine-tuning work in bubbles.
\Cref{fig:eval-sources} shows that at a medium and high load, \sys achieves 24\% and 27\% higher throughput, respectively, compared to \textsc{Tea}.
This is because with higher load, there are more abundant bubbles during prefill which \sys effectively leverages.
At low load, where conventional headroom is abundant, leveraging bubbles does not help, and slightly diminishes throughput by 3\%.
We attribute this decrease to the additional overhead required by the \sys scheduling mechanism.

\subsection{How does \sys protect serving latency?}
\label{sec:eval-latency}

\begin{figure}[t]
  \centering
  \inputplot{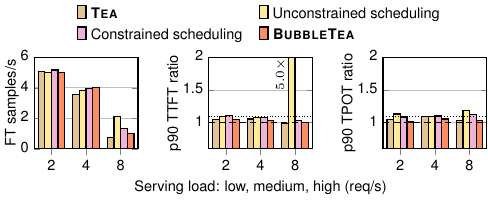}
  \caption{Fine-tuning throughput (left), p90 \ac{TTFT} (middle) and p90 \ac{TPOT} (right) relative to serving alone, for \textsc{Tea}, \emph{unconstrained scheduling}, \emph{constrained scheduling}, and \sys on \textsc{Qwen3-30B-A3B} with 4k prompts across serving loads. The dashed line marks serving alone and the dotted line a 10\% increase.}
  \label{fig:eval-latency-ablation}
\end{figure}

The \sys scheduler should not increase inference latency.
To better understand how our scheduler protects this latency, we progressively enable subcomponents of our scheduler and report in \Cref{fig:eval-latency-ablation} the fine-tuning throughput (left), p90 \ac{TTFT} (middle) and p90 \ac{TPOT} (right) for varying serving loads and scheduler configurations, when using prompts with length 4k.
We highlight the results of prompt lengths of 4k because the effects of these different optimizations are most pronounced at this prompt length, making it easiest to distinguish their individual contributions.

\emph{Unconstrained scheduling} admits fine-tuning work on every GPU during idle time until that GPU has completed the all-reduce operation.
This maximizes the amount of training exposed inside prefill, but can compete for compute needed by serving and can run training on the most-loaded GPU that determines the layer's critical path.

 \emph{Constrained scheduling} adds four mechanisms that protect serving latency. First, waiting for admission becomes a stream-ordered device wait that consumes no GPU compute.
Second, the two fine-tuning chains run concurrently only while prefill takes less than 70\% of recent serving time and are serialized otherwise (\S\ref{sec:design-chunk}).
Third, prefills shorter than 4k tokens are skipped because their bubbles are too small to justify admission.
Fourth, each GPU stops admitting new chunks a margin $m$ before its bubble ends (\emph{early close}), so that no running chunk delays the all-reduce (\S\ref{sec:design-optC}).

\Cref{fig:eval-latency-ablation} shows that at a high load, \emph{unconstrained scheduling} increases fine-tuning throughput from 0.73 to 2.11 samples/s compared to \textsc{Tea}, but p90 \ac{TTFT} increases to 5.03$\times$ serving alone and p90 \ac{TPOT} to 1.19$\times$.
Fine-tuning that reaches the serving-critical GPU extends the prefill and can queue subsequent requests.
Exposing intra-layer capacity is therefore not sufficient; admission must avoid interfering with the critical path.
\emph{Constrained scheduling} reduces p90 \ac{TTFT} to 1.03$\times$ and p90 \ac{TPOT} to 1.13$\times$, while retaining 1.33 samples/s.
Therefore, \emph{our scheduling constraints are key to protect serving latency.}

Constrained scheduling, however, can still admit fine-tuning work on the most-loaded GPU because it does not distinguish GPUs by their predicted slack.
\sys uses the router-derived budget to prevent that GPU from admitting work and to stop other GPUs before their predicted bubbles end.
\Cref{fig:eval-latency-ablation} shows that this reduces p90 \ac{TTFT} and \ac{TPOT} further to 1.01$\times$ and
1.04$\times$ compared to serving only, respectively, at 1.00 samples/s of fine-tuning throughput.

\begin{figure}[t]
  \centering
  \inputplot{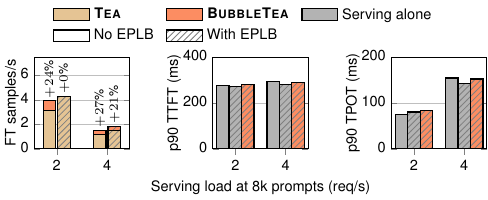}
  \caption{Fine-tuning throughput (left), p90 \ac{TTFT} (middle) and p90 \ac{TPOT} (right) on \textsc{Qwen3-30B-A3B} with prompts of 8k length at medium (2 requests/s) and high (4 requests/s) serving load. Left: each bar shows the throughput of \textsc{Tea}, with the additional throughput of \sys from bubble filling stacked on top; labels give the gain from bubble filling. Middle and right: latency of serving alone without and with \ac{EPLB}, and of  \sys with \ac{EPLB}.}
  \label{fig:eval-eplb}
\end{figure}

\subsection{Composition with load balancing optimizations}
\label{sec:eval-eplb}

We next ask whether \sys remains useful when serving already uses an expert load-balancing optimization.
We use \vllm's \ac{EPLB}, a load-balancing  mechanism that periodically replicates and re-places experts across GPUs based on their observed load, so that each GPU receives a more even share of expert work.
We enable it on \textsc{Qwen3-30B-A3B} with 16 redundant experts.
Across our workloads, \ac{EPLB} reduces the fraction of prefill spent in bubbles from 11--17\% to 8--13\%.
Because \ac{EPLB} also changes how much decode and idle-time headroom serving leaves, we break down the contribution of placing fine-tuning in decode and idle time with \textsc{Tea}, and in the bubbles with \sys.

\Cref{fig:eval-eplb} shows the results for 8k-token prompts at medium and high load.
At high load and across prompt lengths, bubble filling increases fine-tuning throughput over \textsc{Tea} by 12--23\% with \ac{EPLB}, compared to 27--37\% without it.
This is because \ac{EPLB} balances expert load averaged over recent steps, while the routing of each individual prefill still deviates from that average, so some GPUs still finish their experts early in every layer.
At high load, where decode and idle-time headroom is scarce, these residual bubbles are the main remaining source of fine-tuning capacity.
p90 \ac{TTFT} and \ac{TPOT} stay within 7\% of serving alone.

At medium load, bubble filling changes fine-tuning throughput by $-2$ to $+0.5$\%.
As expected, decode and idle-time headroom is plentiful at this load, and \textsc{Tea} already uses it.
The only workload where latency exceeds 7\% is also at medium load: at 16k-token prompts and 1 request/s, p90 \ac{TPOT} reaches 1.32$\times$ serving alone (Appendix~\ref{sec:appx-eplb}).
This suggests \sys should leave residual bubbles unused when conventional headroom already suffices.

Overall, \ac{EPLB} and bubble filling are complementary: \ac{EPLB} shrinks bubbles but cannot remove the imbalance within each prefill, and \sys turns what remains into 12--23\% more fine-tuning throughput at high load while protecting serving latency.
Bubble filling therefore pays off most when serving is busiest, which is exactly when conventional decode and idle-time headroom runs out.

\section{Related Work}
\label{sec:related}

\paragraph{MoE routing and serving optimizations.}
Load imbalance in sparse \ac{MoE} execution is well known and has been addressed through routing objectives and capacity constraints~\cite{switch,gshard}, expert replication and placement, and
serving-time mechanisms such as \textsc{EPLB}~\cite{eplb}, \textsc{LPLB}~\cite{lplb}, and \textsc{Waterfill}~\cite{waterfill}.
\textsc{DeepEP}~\cite{deepep}, \textsc{pplx}-kernels~\cite{pplxkernels}, and lower-precision execution further reduce communication and expert compute.
\sys is complementary: it uses the residual idle capacity that remains after these optimizations are applied.

\paragraph{Inference--fine-tuning colocation and GPU multiplexing.}
\flexllm~\cite{flexllm}, \llmstation~\cite{llmstation},
\harli~\cite{harli}, and \deltaserve~\cite{deltaserve} colocate latency-critical inference with best-effort fine-tuning by exploiting prefill and decode underutilization or serving-level headroom alongside idle periods.
\sys instead targets per-GPU slack that appears \emph{inside} expert-parallel prefill, requiring admission from the current router state rather than serving-level signals.
More broadly, GPU- and LLM-multiplexing systems use temporal sharing, operator interleaving, spatial partitioning, or model-residency management~\cite{gandiva,antman,orion,nvidiamps,nvidiamig,
muxserve,qlm,serverlessllm,aegaeon,prism}.
These systems allocate resources across workloads or models, whereas \sys reclaims per-GPU slack inside an ongoing \ac{MoE} forward pass.
\textsc{ConServe}~\cite{conserve} co-serves online and offline inference of one model and argues against lending idle cycles to training because of process- or container-switch costs; \sys avoids that cost by training on the served weights in place.

\paragraph{Multi-tenant \ac{LoRA} serving.}
\textsc{Punica}~\cite{punica}, \textsc{S-LoRA}~\cite{slora}, and \textsc{dLoRA}~\cite{dlora}
improve inference across many already-trained \ac{LoRA} adapters sharing one base model.
They optimize adapter \emph{serving}, whereas \sys uses otherwise-idle serving capacity to \emph{train} adapters and produces ordinary checkpoints compatible with such downstream systems.

\section{Conclusion}
\label{sec:conclusion}

Expert-parallel \ac{MoE} serving leaves GPUs idle in a way dense serving does not: routing skew makes lightly loaded GPUs wait for the most-loaded one at every layer, and these \emph{bubbles} persist even after applying load balancing optimizations.
\sys predicts each GPU's bubble from the router's decision, admits fine-tuning only inside it, and fills it with
compiler-derived chunks with fine-tuning work.

\bibliographystyle{mlsys2025-arxiv}
\bibliography{main.bib}

\section*{AI Usage Disclosure}

We used large language model assistants to help
write this paper: to improve wording and spell cheking, to generate and edit plots, to check claims and
numbers against our data and the cited works, to search the literature,
and to verify bibliographic metadata. AI was used to implement several parts of the system, write and run scripts for the experiments and to generate the plots.
All AI-assisted content was reviewed and edited by the authors, who take
full responsibility for the paper.

\appendix
\clearpage

\appendix

\setcounter{topnumber}{4}
\setcounter{bottomnumber}{2}
\setcounter{totalnumber}{6}
\setcounter{dbltopnumber}{3}
\renewcommand{\topfraction}{0.95}
\renewcommand{\bottomfraction}{0.6}
\renewcommand{\dbltopfraction}{0.95}
\renewcommand{\textfraction}{0.05}
\renewcommand{\floatpagefraction}{0.75}
\renewcommand{\dblfloatpagefraction}{0.75}
\makeatletter
\setlength{\@fptop}{0pt}\setlength{\@fpbot}{0pt plus 1fil}
\setlength{\@dblfptop}{0pt}\setlength{\@dblfpbot}{0pt plus 1fil}
\setlength{\@fpsep}{14pt plus 4pt}\setlength{\@dblfpsep}{14pt plus 4pt}
\makeatother

\section{Characterization details}
\label{sec:appx-char}

\subsection{Dense and expert-parallel MoE layers}
\label{sec:appx-arch}

\begin{figure}[t]
  \centering
  \inputplot{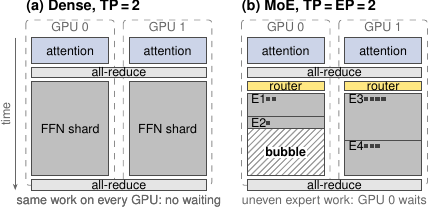}
  \caption{\textbf{Where expert-parallel bubbles come from.} Time runs
    downward within each GPU. (a)~A dense layer under \ac{TP} splits
    work evenly, so no GPU waits. (b)~An \ac{MoE} layer under
    \ac{TP}$=$\ac{EP} places different experts on different GPUs; the
    router sends GPU~0's experts fewer tokens (squares), so GPU~0
    finishes early and idles until GPU~1 reaches the all-reduce that
    combines the expert outputs. Schematic; sizes are illustrative.}
  \label{fig:appx-arch}
\end{figure}

Figure~\ref{fig:appx-arch} contrasts a dense Transformer layer with an
\ac{MoE} layer on two GPUs. Under \ac{TP}, a dense layer shards
attention heads and feed-forward weights evenly, so every GPU performs
the same work on all tokens and reaches each all-reduce at the same
time. In the \ac{MoE} layer, attention is still sharded by \ac{TP}, and
every GPU computes the same routing decision, but \ac{EP} places whole
experts on different GPUs. Each GPU then runs only the tokens routed to
its local experts. When the router sends more tokens to one GPU's
experts, the other GPU finishes early and waits at the all-reduce that
combines the expert outputs: this wait is the bubble of
Figure~\ref{fig:anatomy}.

\subsection{Bubble size across models}
\label{sec:appx-context}

\begin{figure}[t]
  \centering
   \inputplot{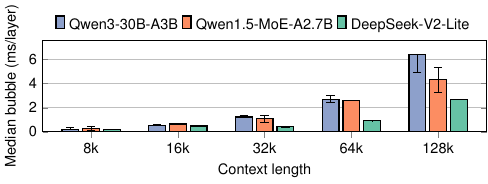}
  \caption{
    \textbf{Per-layer bubble duration grows with prefill size.}
    Median bubble across three \ac{MoE} architectures under BF16
    execution without load balancing. Error bars show min/max across
    repetitions.
  }
  \label{fig:context}
\end{figure}

Figure~\ref{fig:context} shows how the per-layer bubble grows with
prefill size for the three \ac{MoE} models of \S\ref{sec:char-context},
in the default setup of \S\ref{sec:characterization}. From 8k to 128k
tokens, the median bubble grows from 0.22 to 6.41\,ms/layer on
Qwen3-30B-A3B, from 0.28 to 4.34\,ms/layer on Qwen1.5-MoE-A2.7B, and
from 0.20 to 2.70\,ms/layer on DeepSeek-V2-Lite.

\subsection{Lower-precision execution}
\label{sec:appx-fp8}

\begin{figure}[t]
  \centering
  \inputplot{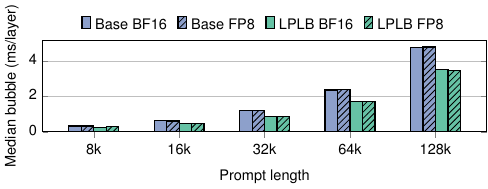}
  \caption{\textbf{Lower precision leaves the bubble unchanged.}
    Median per-layer bubble under BF16 and FP8 (hatched), without load
    balancing (Base) and with LPLB, in the setup of
    Figure~\ref{fig:lb-sweep}: DeepSeek-V2-Lite, SGLang, 8$\times$H100,
    \ac{EP}$=8$. Context denotes aggregate tokens across the 8-way
    \ac{EP} group.}
  \label{fig:appx-fp8}
\end{figure}

Lower-precision execution attacks a different part of the problem.
FP8 reduces the cost of expert computation, but does not change how
the router distributes expert work across GPUs. Nor, in our measured
regime, does it reduce expert computation enough to remove the
compute saturation during prefill: additional routed work on one GPU
still extends its completion time. Figure~\ref{fig:appx-fp8} compares
BF16 and FP8 in the load-balancing setup of Figure~\ref{fig:lb-sweep}
(DeepSeek-V2-Lite, SGLang, 8$\times$H100, \ac{EP}$=8$). At 128k,
moving from BF16 to FP8 changes the bubble from 4.79 to 4.81\,ms/layer
without load balancing and from 3.53 to 3.50\,ms/layer with LPLB, and
it changes by at most 0.05\,ms/layer at every other context. Faster
expert execution therefore does not, by itself, remove the GPU-to-GPU
timing differences caused by uneven routed work.

\subsection{Bubble predictability}
\label{sec:appx-predict}

\begin{figure}[t]
  \centering
  \inputplot{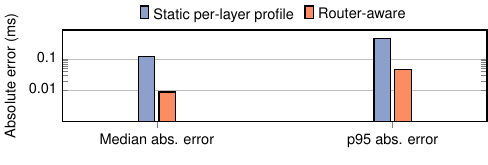}
  \caption{\textbf{The router's assignment predicts each GPU's bubble.}
    Absolute error of the router-aware predictor and of a static
    per-layer profile on Qwen3-30B-A3B, 4$\times$A100,
    \ac{TP}$=$\ac{EP}$=4$, pooled over five serving loads.}
  \label{fig:appx-predict}
\end{figure}

We measure how well the router's token assignment predicts each GPU's
bubble on Qwen3-30B-A3B, served by \vllm on 4$\times$A100 with
\ac{TP}$=$\ac{EP}$=4$ and no expert load balancing, at five loads
(4k tokens at 8 requests/s, 8k at 4 and 2, 12k at 2, and 2k at 1;
400 \textsc{arXiv} requests each). For every prefill layer and GPU
(\num{193920} samples), we compare the measured bubble with two
predictors: the deployed router-aware predictor
(\S\ref{sec:design-predict}), and a static profile that assigns every
GPU the mean bubble of that layer from the first half of the run. Both
are evaluated on the second half. \Cref{fig:appx-predict} shows that
the router-aware predictor reduces the median absolute error from
\SI{123}{\micro\second} to \SI{8.9}{\micro\second} and the p95 from
\SI{452}{\micro\second} to \SI{45}{\micro\second}. With \ac{EPLB}
enabled, its median error is \SI{9.5}{\micro\second} (p95
\SI{43}{\micro\second}). Our probe resolves durations to
\SIrange{4}{8}{\micro\second}, so the router-aware median error is at
the measurement resolution.

\section{Colocation systems and the planner experiment}
\label{sec:appx-gap}

Colocation runs best-effort work on resources that latency-critical
serving leaves unused while preserving its \acp{SLO}
~\cite{flexllm,llmstation,harli,deltaserve}. Fine-tuning is a natural
candidate because it is throughput-oriented and delay-tolerant, and
\ac{PEFT} methods such as \ac{LoRA}~\cite{lora} train only small
adapters while reusing the serving model's resident weights. Because
GPU kernels are difficult to preempt once launched~\cite{nonpreempt},
colocation systems must both identify spare capacity and expose work
units short enough to fit inside it.

Existing inference--fine-tuning systems already decompose training for
this purpose, but make admission decisions at serving boundaries using
signals such as batch composition, iteration timing, or engine
headroom (\Cref{tab:colocation_systems}). Expert-parallel bubbles have a different
granularity: they open and close inside one \ac{MoE} layer and differ
across GPUs. Using them therefore requires both \emph{bubble-sized
work} and a \emph{per-GPU} admission decision that reveals where slack
exists and for how long. The current router assignment provides this missing signal
(\S\ref{sec:design-predict}).

\begin{table*}[t]
\centering
\small
\caption{
\textbf{Existing colocation systems reason about serving-level headroom, whereas \sys{} identifies slack per GPU inside an \ac{MoE} layer.}
}
\label{tab:colocation_systems}
\begin{tabular}{lllll}
\toprule
\emph{System} & \emph{Capacity targeted} & \emph{Work unit} & \emph{Admission scope} & \emph{Signal} \\
\midrule
\flexllm~\cite{flexllm}
& serving headroom
& token slot
& serving iteration
& batch composition \\
\llmstation~\cite{llmstation}
& decode / idle headroom
& tasklet
& serving iteration
& iteration timing \\
\harli~\cite{harli}
& decode underutilization
& SM partition
& serving iteration
& predicted latency \\
\deltaserve~\cite{deltaserve}
& serving / idle headroom
& LoRA batch
& serving iteration
& predicted latency \\
\midrule
\sys{}
& \ac{EP} bubble
& fine-tuning chunk
& GPU $\times$ \ac{MoE} layer

& router top-$k$ assignment \\
\bottomrule
\end{tabular}
\end{table*}

We first isolate the planning requirement. We replay a
Qwen3-30B-A3B serving trace (\ac{TP}$=$\ac{EP}$=$2, 128k-token
prefill) against the same 486 measured fine-tuning chunks
(median 567\,$\mu$s) while varying only the information available to
the planner. A \emph{router-blind} planner assigns every GPU a
historical layer-level slack estimate; an \emph{oracle GPU-agnostic}
planner instead receives the true mean bubble of the current layer but
still assigns the same budget to every GPU; and a \emph{router-aware}
planner estimates slack separately for each GPU from the current
routing decision. We compare their \emph{bubble utilization}, the
fraction of available bubble GPU-time occupied by training, at
approximately the same 1\% overrun rate.

\begin{table}[t]
  \centering
  \small
  \caption{
    \textbf{Per-GPU planning is required to use bubbles.}
    All planners use the same fine-grained fine-tuning chunks and operate
    near a 1\% overrun rate.}

  \label{tab:gap-planner}
  \begin{tabular}{lrr}
    \toprule
    Planner & Bubble utilization & Overrun rate \\
    \midrule
    Router-blind
      & 1.28--1.31\% & 1.04\% \\
    Oracle GPU-agnostic
      & 1.22--1.30\% & 1.04\% \\
    Router-aware
      & 42.74--45.14\% & 0.83--1.04\% \\
    \bottomrule
  \end{tabular}
\end{table}

Table~\ref{tab:gap-planner} shows that router-aware planning recovers
42.7--45.1\% of bubble capacity, compared with about 1.3\% for either
GPU-agnostic planner. Even knowing the exact \emph{average} slack of
the current layer is therefore insufficient: the planner must know
which GPU owns that slack.

Shorter work units determine how much of this per-GPU capacity can be
used. With router-aware planning, reducing the median chunk duration
from 4.13\,ms (97 chunks) to 567\,$\mu$s (486 chunks) raises bubble
utilization from 37.95--38.46\% to 42.74--45.14\%. Yet the
router-blind planner remains near 1--1.5\% under both partitions.
Thus, finer execution helps only after the planner identifies the
right GPU: chunk size determines \emph{how much} work fits, while
router state determines \emph{where} it can safely run.

\section{Experimental setup details}
\label{sec:appx-eval}

\subsection{Experimental setup for \Cref{sec:char-obs}}
\label{sec:appx-setup-obs}

\paragraph{Observation 1.}
Routing skew and decode bubbles use Qwen3-30B-A3B on vLLM with
\ac{TP}$=$\ac{EP}$=2$ on two A100 GPUs, eager mode, and no load
balancing. Skew is the most-loaded GPU's routed tokens divided by the
mean, summed over all \ac{MoE} layers, from vLLM's EPLB balancedness
logging with no redundant experts and no rebalancing. Each dataset
sends 30 requests one at a time, so each request contributes one
prefill step and one decode step per generated token. Decode bubbles
are measured from GPU kernel timestamps recorded with
\texttt{torch.profiler}.

\paragraph{Observation 2.}
Load-balancing results (\Cref{fig:lb-sweep}) use DeepSeek-V2-Lite on
SGLang with 8$\times$H100, \ac{EP}$=8$, BF16, and \textsc{arXiv}
prompts. Each point is the median over 1,248 samples: six prefill
forwards, 26 \ac{MoE} layers, and eight GPUs.

\paragraph{Observation 3.}
The colocation systems and the planner experiment are described in
Appendix~\ref{sec:appx-gap}.

\subsection{Evaluation setup}
\label{sec:appx-setup}

This section gives the full configuration summarized in \S\ref{sec:eval-method}.

\paragraph{Hardware and software.}
All experiments run on one node with four NVIDIA A100-SXM4-80GB GPUs
connected by NVLink. Unless stated otherwise, serving uses vLLM with
\ac{TP}=\ac{EP}=4, BF16 weights, eager execution, chunked prefill and
prefix caching disabled, and a 0.85 GPU-memory budget. \sys{} runs one
fine-tuning worker per GPU under CUDA MPS, as described in
\S\ref{sec:design}, with a 50\,$\mu$s chunk budget and an early-close
margin of $m=60\,\mu$s (\S\ref{sec:design-optC}).

\paragraph{Models.}
We evaluate three MoE architectures with different expert
organizations. Qwen3-30B-A3B (48 layers, 128 experts, top-8) uses only
routed experts, while Qwen1.5-MoE-A2.7B (24 layers, 60 routed experts,
top-4) and DeepSeek-V2-Lite (27 layers, 64 routed experts, top-6) also
contain shared experts. Qwen1.5-MoE's
configuration declares an 8k context; for longer prompts we extend its
rotary cache to the served length identically for every system.

\paragraph{Serving workloads and metrics.}
Our controlled workloads use arXiv summarization prompts truncated to
an exact input length between 1k and 32k tokens, each generating 32
output tokens. Short outputs make the workload prefill-dominated, where
bubbles arise, and resemble short-form tasks such as extraction or
classification; \S\ref{sec:eval-e2e} also replays production traces.

For every model and prompt length, we choose three arrival rates:
\emph{high} is just below the point at which serving alone begins
sustained queueing, while \emph{medium} and \emph{low} are fixed
fractions of that rate. Each steady-state run serves 400 requests and
is repeated thrice unless stated otherwise. Latency statistics discard
the first and last 10\% of requests and pool the remaining requests
across repetitions. We report fine-tuning throughput as completed
samples/s while requests are being served, together with tail
\ac{TTFT} and \ac{TPOT} relative to the corresponding serving-only
execution.

\paragraph{Fine-tuning workload.}
Every system trains the same job: LoRA rank 16 on the attention
projections, using alpaca-cleaned samples padded to 128 tokens and
AdamW with learning rate $2\cdot10^{-4}$. We adapt $(q,k,v,o)$ for
Qwen3-30B-A3B and Qwen1.5-MoE-A2.7B, and
$(q,kv_a,kv_b,o)$ for DeepSeek-V2-Lite.

\paragraph{Baselines.} We compare four systems.
\begin{itemize}
  \item \textbf{Serving only} provides the inference-latency reference.

  \item \textbf{\Uncoord} runs a stock torchtune LoRA trainer
  using FSDP alongside vLLM under CUDA MPS. Because its interference
  depends on the amount of GPU compute available to the trainer, we
  run it with the trainer's SM share capped at 6\% and 40\% and
  uncapped. The main text shows the 40\% cap;
  Appendix~\ref{sec:appx-naive} compares all three.

  \item \textbf{LLMStation}~\cite{llmstation} represents iteration-level colocation through suspendable tasklets. We update its implementation to the current vLLM (v1 engine) and PyTorch versions, and extend its training forward to Qwen3-MoE with shared serving weights. Its released fusion engine assumes dense projections with fixed weight placement, which does not hold for expert-parallel \ac{MoE}, especially with EPLB. We therefore use its separate PyTorch training path alongside inference under CUDA MPS with EPLB disabled. To obtain a representative baseline, we evaluated several expert-execution paths including a naive per-expert forward, reuse of \sys{}'s compiler-generated chunks and a packed-GEMM implementation. Executing padded batched GEMMs through cuBLAS gave the best LLMStation performance. This path does not use \sys{}'s compiler generated chunks or A100-specific grouped-GEMM kernel. LLMStation's latency ratios are measured relative to the same engine with training disabled, so differences between serving engines do not count against either system.

  \item \textbf{\sys{}} is the full system. It combines conventional
  decode/idle-time colocation with router-aware intra-layer bubble
  filling. Unless stated otherwise, it uses all mechanisms of
  \S\ref{sec:design}, including device-side readiness, paired chains,
  and early close.
\end{itemize}

We do not compare against colocation systems that use compute headroom
during prefill: prefill saturates the GPU at a few hundred
tokens~\cite{sarathi}, so at our prompt lengths that headroom is gone,
and the remaining serving-level opportunities, idle and decode time, are
the ones LLMStation targets.

\section{Additional experiments}
\label{sec:appx-experiments}

This section extends the evaluation of \S\ref{sec:evaluation} and
follows the order of its subsections.
For the end-to-end comparison (\S\ref{sec:eval-e2e}),
\S\ref{sec:appx-e2e} reports all three models and 1k--32k prompts,
\S\ref{sec:appx-tp2} repeats the Qwen3-30B-A3B comparison on two GPUs,
\S\ref{sec:appx-traces} describes both production traces and their full
results, and \S\ref{sec:appx-naive} sweeps the \ac{SM} cap of \uncoord.
For the sources of throughput and the latency protections
(\S\ref{sec:eval-ablation} and \S\ref{sec:eval-latency}),
\S\ref{sec:appx-ablation} repeats both breakdowns at other prompt
lengths, \S\ref{sec:appx-margin} selects the early-close margin,
\S\ref{sec:appx-threshold} tests the 4k-token bubble-filling threshold,
\S\ref{sec:appx-capacity} measures the decode and idle-time headroom
that serving leaves at each load, and \S\ref{sec:appx-bubble-use} shows
where the bubble capacity that \sys{} leaves unused goes.
\S\ref{sec:appx-eplb} extends the composition with \ac{EPLB}
(\S\ref{sec:eval-eplb}) to all workloads. Finally,
\S\ref{sec:eval-overheads} isolates the overhead of the \sys{} runtime
and verifies that chunked execution preserves training behavior.

\begin{figure*}[t]
  \centering
  \inputplot{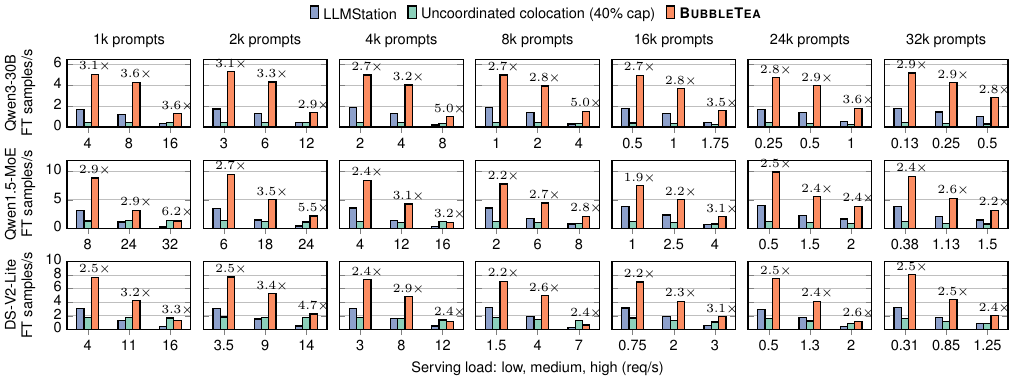}
  \caption{\textbf{Fine-tuning throughput while serving, all models.}
    Samples/s for \sys{}, LLMStation, and \uncoord (40\% cap)
    across three \ac{MoE} models, prompt lengths, and serving loads.
    Labels report \sys{}'s throughput relative to LLMStation.}
  \label{fig:appx-e2e-tput}
\end{figure*}

\begin{figure*}[t]
  \centering
  \inputplot{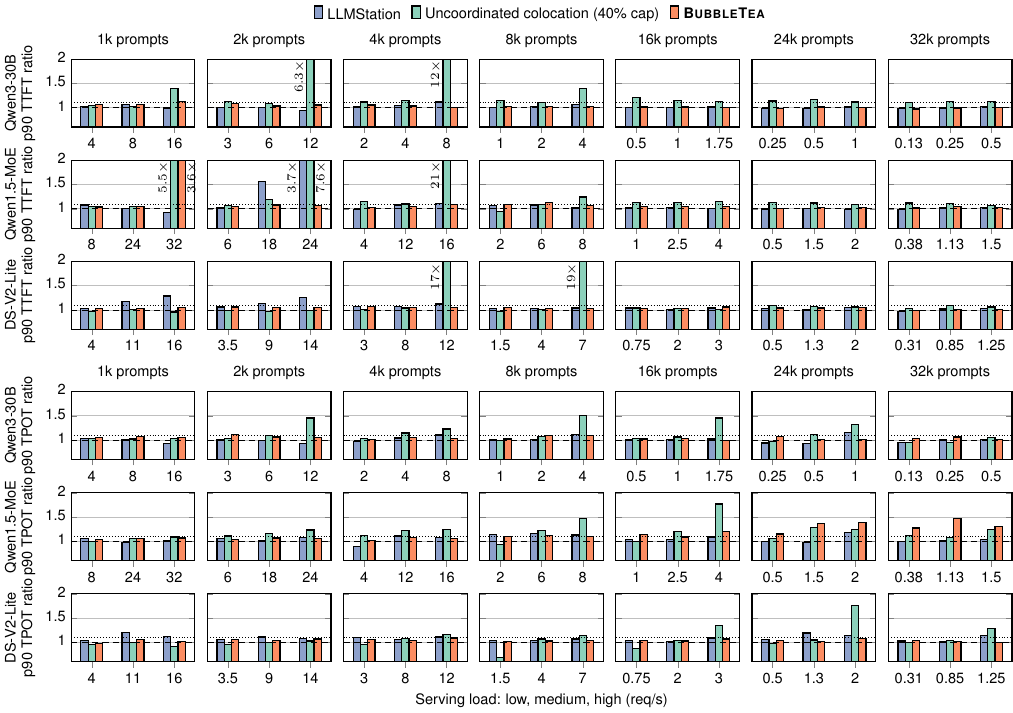}
  \caption{\textbf{Serving latency under colocation, all models.}
    p90 \ac{TTFT} (top) and p90 \ac{TPOT} (bottom), relative to the
    corresponding serving-only execution. The dashed line marks serving
    alone and the dotted line a 10\% increase. Bars above 2$\times$ are
    clipped, with their value printed beside them. \Uncoord uses
    the 40\% cap.}
  \label{fig:appx-e2e-lat}
\end{figure*}

\subsection{End-to-end results for all models}
\label{sec:appx-e2e}

\S\ref{sec:eval-e2e} compares \sys{} with the baselines on Qwen3-30B-A3B
at 4k--32k prompts. Figures~\ref{fig:appx-e2e-tput} and~\ref{fig:appx-e2e-lat} extend
Figures~\ref{fig:eval-e2e-tput} and~\ref{fig:eval-e2e-lat} to
Qwen1.5-MoE-A2.7B and DeepSeek-V2-Lite, and to 1k, 2k, 8k, and 24k prompts.
Both models follow the same trends as Qwen3-30B-A3B: across all three models and 1k--32k prompts, \sys{} trains 1.9--6.2$\times$ faster than \llmstation{} (median 2.8$\times$). At 1k and 2k
tokens, prefills are too short for \sys{} to open useful windows,
so its gains there come from its runtime mechanisms on decode and
idle time (\S\ref{sec:eval-ablation}).

\sys{} keeps both p90 \ac{TTFT} and \ac{TPOT} within 10\% of serving
alone in 50 of the 63 workloads. Of the other 13, eight are the main
exception, \ac{TPOT} on Qwen1.5-MoE-A2.7B at long prompts: from 16k to
32k tokens, \sys{}'s p90 \ac{TPOT} reaches 1.14--1.47$\times$ serving
alone in every workload but 16k at medium load (1.09$\times$), highest
at 32k and medium load, while \ac{TTFT} stays within 6\%. Its short
decode iterations overlap with the fine-tuning worker's cross-GPU exchanges and
increase \ac{TPOT}. Three more exceed the bound by at most three
points: Qwen1.5-MoE-A2.7B at 8k and medium load (p90 \ac{TTFT}
1.13$\times$, \ac{TPOT} 1.12$\times$, where LLMStation's \ac{TPOT} is
1.15$\times$) and high load (\ac{TPOT} 1.10$\times$), and Qwen3-30B-A3B
at 2k and low load (\ac{TPOT} 1.11$\times$).

The last two are \ac{TTFT} at 1k-token prompts and high load:
Qwen1.5-MoE-A2.7B at 32 requests/s reaches 3.61$\times$ serving alone
and Qwen3-30B-A3B at 16 requests/s 1.13$\times$, while \ac{TPOT} stays
within 7\%. These prompts are below the 4k-token threshold, so \sys{}
trains only in decode and idle time, and the server is already near
saturation: on Qwen1.5-MoE-A2.7B each serving step takes about 29\,ms,
so 32 requests/s keep the GPU about 93\% busy, and serving alone
completes only 29 requests/s. Decode and idle-time training leaves the
median step unchanged but lengthens the slowest ones (p90 from 30 to
34\,ms in one run), and at this utilization that builds a request queue.
This is a limitation of the current policy: \sys{} does not pause
decode- and idle-time training while requests are queued, and we leave
such a queue-aware cutoff to future work.

\subsection{Smaller expert-parallel degree}
\label{sec:appx-tp2}

\S\ref{sec:eval-e2e} uses four GPUs; we test whether \sys{}'s advantage
holds with fewer expert-parallel GPUs, where each GPU hosts more
experts. Table~\ref{tab:appx-tp2} repeats the Qwen3-30B-A3B comparison with
half the GPUs, at \ac{TP}$=$\ac{EP}$=$2. High load is half the
\ac{TP}$=$\ac{EP}$=$4 high load of each prompt length, and medium load about a
quarter of it. \sys{} trains 3.3--3.9$\times$ faster than LLMStation
and keeps both p90 \ac{TTFT} and \ac{TPOT} within 10\% of serving alone
in every workload; the highest is p90 \ac{TPOT} at 8k tokens and high
load (1.10$\times$, against 1.11$\times$ for LLMStation).

\begin{table}[t]
  \centering
  \footnotesize
  \setlength{\tabcolsep}{1.6pt}
  \caption{\textbf{\ac{TP}$=$\ac{EP}$=$2.} Qwen3-30B-A3B, two
    repetitions. LLMS: LLMStation; \sysshort: \sys{}. FT: fine-tuning samples/s; TTFT and TPOT: p90 relative
    to serving alone.}
  \label{tab:appx-tp2}
  \begin{tabular}{llrrrrrrrr}
    \toprule
    & & & \multicolumn{2}{c}{FT} & & \multicolumn{2}{c}{LLMS} & \multicolumn{2}{c}{\sysshort} \\
    \cmidrule(lr){4-5}\cmidrule(lr){7-8}\cmidrule(lr){9-10}
    Len. & Load & req/s & LLMS & \sysshort & Ratio & TTFT & TPOT & TTFT & TPOT \\
    \midrule
    4k & med. & 2 & 1.58 & 5.60 & 3.5$\times$ & 1.05 & 1.08 & 1.04 & 1.05 \\
    4k & high & 4 & 0.95 & 3.46 & 3.6$\times$ & 1.02 & 1.01 & 1.02 & 1.00 \\
    8k & med. & 1 & 1.52 & 5.22 & 3.4$\times$ & 1.04 & 1.05 & 1.02 & 1.01 \\
    8k & high & 2 & 0.91 & 3.51 & 3.9$\times$ & 1.03 & 1.11 & 1.02 & 1.10 \\
    16k & med. & 0.5 & 1.46 & 4.86 & 3.3$\times$ & 1.02 & 1.02 & 1.02 & 1.04 \\
    16k & high & 0.875 & 0.90 & 3.18 & 3.5$\times$ & 1.02 & 1.02 & 1.01 & 1.01 \\
    \bottomrule
  \end{tabular}
\end{table}

\subsection{Production traces}
\label{sec:appx-traces}

\S\ref{sec:eval-e2e} summarizes \sys{}'s behavior under bursty production
arrivals, and \Cref{fig:intro-trace} shows one trace; here we describe
both traces and report full results. Constant-rate workloads let us
control prompt length and serving load, but production arrivals are
bursty. We therefore replay two traces from
the Azure LLM inference trace~\cite{splitwise}, \emph{code} and
\emph{conversation}, on Qwen3-30B-A3B, each over a 10-minute window
(minutes 18--28 and 17--27) at its original arrival rate. The traces
give each request's arrival time and its input and output lengths; we
fill prompts with arXiv text of that length. The code trace has 2,183
requests with long prompts and short outputs (median input 1,506 tokens,
p90 5,823; median output 13, p90 52) and very bursty arrivals: 3.9
requests/s on average but 0--14.6 per 10\,s interval, with a
coefficient of variation of inter-arrival times of 10.3. The
conversation trace has 3,716 requests at a steadier 6.2 requests/s
(3.9--8.2 per interval; coefficient of variation 1.0), with shorter
prompts (median 1,024, p90 4,080) and longer outputs (median 111, p90
421). Prompts of at least 4k tokens carry 47\% of prompt tokens in the
code trace and 10\% in the conversation trace.

\begin{figure}[t]
  \centering
  \inputplot{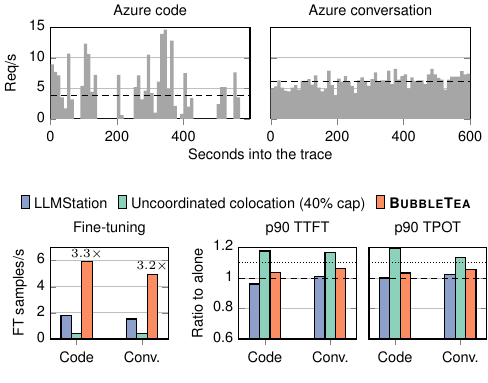}
  \caption{\textbf{Trace-driven serving.}
    Request arrivals (10\,s bins, dashed: mean rate), fine-tuning
    throughput, and p90 serving latency relative to serving alone under
    two production traces. In the latency panels, the dashed line marks
    serving alone and the dotted line a 10\% increase. }
  \label{fig:eval-traces}
\end{figure}

Figure~\ref{fig:eval-traces} shows that \sys{} keeps its advantage
under bursty arrivals: it trains at 5.9 and 5.0 samples/s across the two
traces, compared with 1.8 and 1.5 samples/s for LLMStation, while p90
\ac{TTFT} and \ac{TPOT} remain within 7\% of serving alone.

\subsection{\Uncoord sweep}
\label{sec:appx-naive}

\begin{figure*}[t]
  \centering
  \inputplot{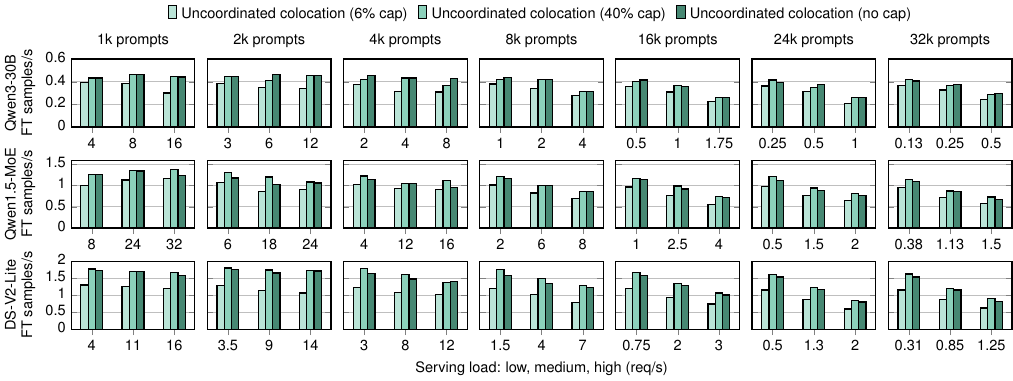}
  \caption{\textbf{\Uncoord sweep: fine-tuning throughput.}
    Samples/s of the torchtune trainer colocated with vLLM, with the
    trainer's \ac{SM} share capped at 6\%, at 40\%, or uncapped.}
  \label{fig:appx-naive-tput}
\end{figure*}

\begin{figure*}[t]
  \centering
  \inputplot{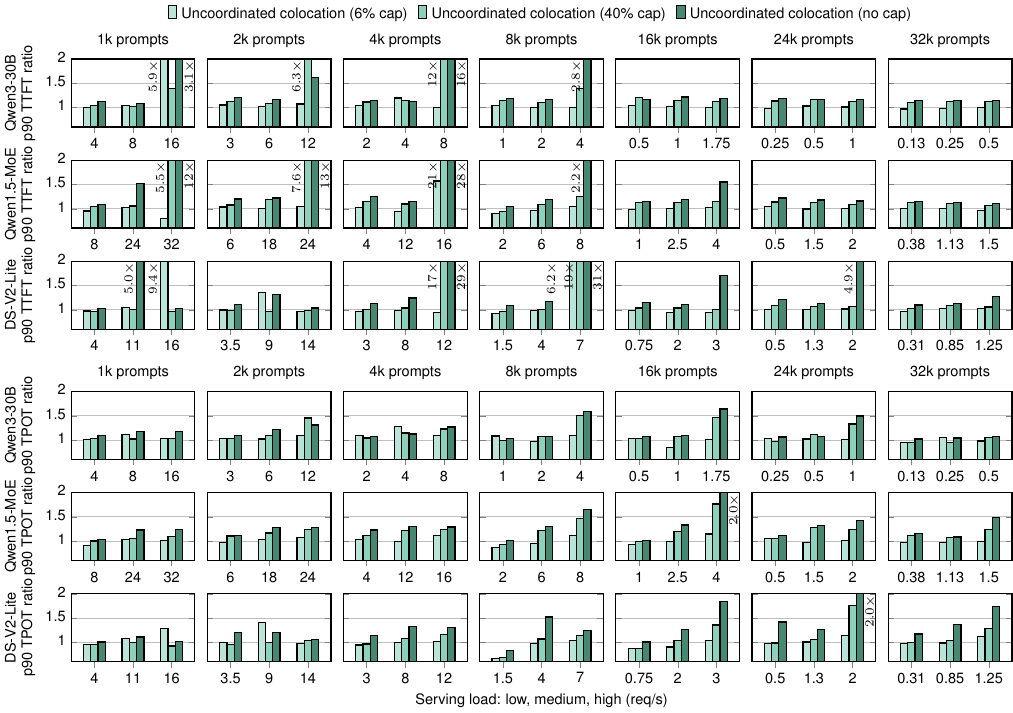}
  \caption{\textbf{\Uncoord sweep: serving latency.}
    p90 \ac{TTFT} (top) and p90 \ac{TPOT} (bottom) relative to serving
    alone, for the three \ac{SM} caps of
    Figure~\ref{fig:appx-naive-tput}. The dashed line marks serving alone
    and the dotted line a 10\% increase. Bars above 2$\times$ are clipped,
    with their value printed beside them.}
  \label{fig:appx-naive-lat}
\end{figure*}

\S\ref{sec:eval-e2e} runs \uncoord with a 40\% \ac{SM} cap and notes that
lifting the cap does not help; here we justify that choice by sweeping
the cap. \Uncoord's interference depends on how much GPU compute its
trainer may use. Figures~\ref{fig:appx-naive-tput}
and~\ref{fig:appx-naive-lat} compare three settings of the trainer's
\ac{SM} cap under CUDA MPS: 6\%, 40\%, and uncapped. Raising the cap
from 6\% to 40\% raises the median fine-tuning throughput across the 63
workloads from 0.80 to 1.05 samples/s, but the number of workloads that
keep both p90 \ac{TTFT} and p90 \ac{TPOT} within 10\% of serving alone
falls from 51 of 63 to 22 of 63. Without a cap, median throughput does
not rise further (1.00 samples/s), and only 6 of 63 workloads stay
within that bound. In the median workload, \sys{} trains 5.9$\times$
faster than \uncoord at the 6\% cap and 4.4$\times$ faster than
at the 40\% cap. In a few high-load workloads \uncoord trains
faster than \sys{}, almost always by inflating p90 \ac{TTFT} by
5--31$\times$. The exception is DeepSeek-V2-Lite at 1k tokens and high
load, where \uncoord at the 40\% cap or uncapped trains about
1.2$\times$ faster than \sys{} while staying within 3\% of serving alone.

\subsection{Ablation details}
\label{sec:appx-ablation}

\begin{figure*}[t]
  \begin{minipage}[t]{\columnwidth}
    \vspace{0pt}\centering
  \inputplot{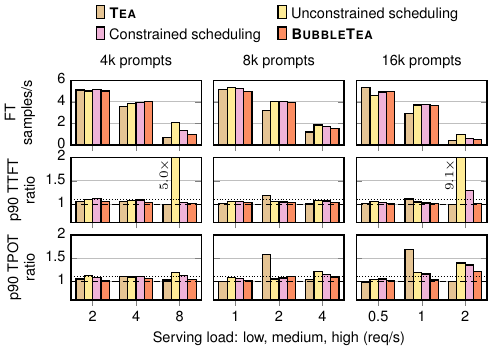}
  \caption{\textbf{Mechanisms that protect serving latency, 4k--16k
    prompts.} Rows: fine-tuning throughput, p90 \ac{TTFT} and p90
    \ac{TPOT} relative to serving alone. Configurations as in
    Figure~\ref{fig:eval-latency-ablation}.}
  \label{fig:appx-latency-ablation}
  \end{minipage}\hfill
  \begin{minipage}[t]{\columnwidth}
    \vspace{0pt}\centering
  \inputplot{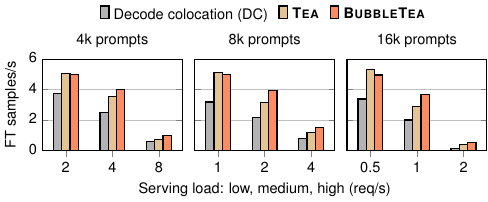}
  \caption{\textbf{Sources of \sys{}'s fine-tuning throughput, 4k--16k
    prompts.} Configurations as in Figure~\ref{fig:eval-sources}.}
  \label{fig:appx-sources}
  \end{minipage}
\end{figure*}

\S\ref{sec:eval-ablation} and \S\ref{sec:eval-latency} break down \sys{}'s
throughput and its latency protections at 8k and 4k prompts; here we
check that both hold at other prompt lengths.
Figure~\ref{fig:appx-sources} extends Figure~\ref{fig:eval-sources} to
4k and 16k prompts. The pattern is the same at every length:
\cfgOptColo raises throughput over \cfgColo by 36--60\% at low and
medium load, bubble filling changes throughput by $-7$ to $-2$\% at
low load and adds 13--37\% at medium and high load. At 16k and high load
all three configurations train at most 0.53 samples/s, so the
relative gain of \cfgOptColo there (+159\%) is large but small in
absolute terms.

Figure~\ref{fig:appx-latency-ablation} extends
Figure~\ref{fig:eval-latency-ablation} to 8k and 16k prompts. Unthrottled
bubbles hurt serving only near saturation, and most at 16k prompts and
2 requests/s, where p90 \ac{TTFT} reaches 9.15$\times$ serving alone.
At 8k and 4 requests/s the server keeps enough headroom that the same
configuration keeps p90 \ac{TTFT} within 1.08$\times$, but its p90
\ac{TPOT} reaches 1.22$\times$. At 16k and high load, \sys{}
keeps \ac{TTFT} at 1.01$\times$ but its p90 \ac{TPOT} is 1.21$\times$.

\subsection{Sensitivity to the early-close margin}
\label{sec:appx-margin}

In \S\ref{sec:eval-latency}, the early close is one of the protections
that keep training off the serving critical path; this subsection
selects its margin. \sys{} closes each admission window a margin $m$ before its predicted
end (\S\ref{sec:design-optC}). We select $m$ with the sweep in
Table~\ref{tab:appx-margin}, which varies $m$ from 0 to 120\,$\mu$s on
Qwen3-30B-A3B at three high-load workloads, where bubble filling
matters most (\S\ref{sec:eval-ablation}); all other settings are those
of \sys{}. The mechanism sets a floor on $m$: the margin should cover
one 50\,$\mu$s chunk, so that a chunk admitted just before
the close ends before the predicted end, and the predictor's p95 error
of 45\,$\mu$s (\S\ref{sec:design-predict}). The
sweep separates two regimes. Margins above 60\,$\mu$s cost 10--19\% of
fine-tuning throughput without lowering serving latency: p90 \ac{TTFT}
stays within 6\% of serving alone at every margin, and p90 \ac{TPOT}
does not fall (at 16k it rises to 1.22--1.25$\times$). Between 0 and
60\,$\mu$s, throughput and latency differ by no more than run-to-run
variation. Closing the window at the predicted end is what protects
serving (\S\ref{sec:eval-latency}); the margin absorbs prediction
error. We therefore pick $m=60\,\mu$s, the largest margin that costs no
measurable throughput: it clears the floor and gives the most headroom
against mispredictions at no cost.

\begin{table}[t]
  \centering
  \small
  \setlength{\tabcolsep}{4pt}
  \caption{\textbf{Early-close margin sweep.} Qwen3-30B-A3B,
    \ac{TP}$=$\ac{EP}$=$4. FT: fine-tuning samples/s; TPOT: p90
    \ac{TPOT} relative to serving alone. Two repetitions at 4k, one at
    8k and 16k; at 4k and 60\,$\mu$s, one run with a queueing episode
    (p90 \ac{TTFT} 559\,ms against 192\,ms) is excluded. Bold: the margin
    selected for \sys{}, 60\,$\mu$s.}
  \label{tab:appx-margin}
  \begin{tabular}{rrrrrrr}
    \toprule
    & \multicolumn{2}{c}{4k, 8 req/s} & \multicolumn{2}{c}{8k, 4 req/s} & \multicolumn{2}{c}{16k, 2 req/s} \\
    \cmidrule(lr){2-3}\cmidrule(lr){4-5}\cmidrule(lr){6-7}
    $m$ ($\mu$s) & FT & TPOT & FT & TPOT & FT & TPOT \\
    \midrule
    0 & 0.96 & 1.05 & 1.59 & 1.10 & 0.53 & 1.09 \\
    30 & 0.98 & 1.05 & 1.55 & 1.11 & 0.51 & 1.13 \\
    \textbf{60} & \textbf{0.97} & \textbf{1.05} & \textbf{1.63} & \textbf{1.08} & \textbf{0.51} & \textbf{1.08} \\
    90 & 0.86 & 1.06 & 1.39 & 1.13 & 0.43 & 1.22 \\
    120 & 0.83 & 1.05 & 1.47 & 1.09 & 0.42 & 1.25 \\
    \bottomrule
  \end{tabular}
\end{table}

\subsection{Bubble-filling threshold}
\label{sec:appx-threshold}

\cfgThrottled and \sys{} skip prefills shorter than 4k tokens
(\S\ref{sec:eval-latency}, \S\ref{sec:design-predict}); here we test
whether this threshold is needed. To do so, we rerun
Qwen3-30B-A3B at 2k-token prompts, where it never fires, with bubble
filling enabled in every prefill step. At low load (3 requests/s),
removing the threshold lowers fine-tuning throughput from 5.11 to 4.69
samples/s ($-8$\%), and p90 \ac{TTFT} and \ac{TPOT} rise from
1.07$\times$ and 1.05$\times$ serving alone to 1.13$\times$ and
1.09$\times$. At high load (12 requests/s), throughput falls from 1.57
to 1.39 samples/s ($-11$\%), and p90 \ac{TTFT} and \ac{TPOT} rise from
1.04$\times$ and 1.05$\times$ to 1.08$\times$ and 1.13$\times$. At 2k
tokens the bubbles are too short to amortize prediction and gating, so
filling them only adds contention. Both settings use two
repetitions, with latency relative to serving alone in the same
repetition.

\subsection{Spare capacity while serving}
\label{sec:appx-capacity}

\begin{figure*}[t]
  \centering
  \inputplot{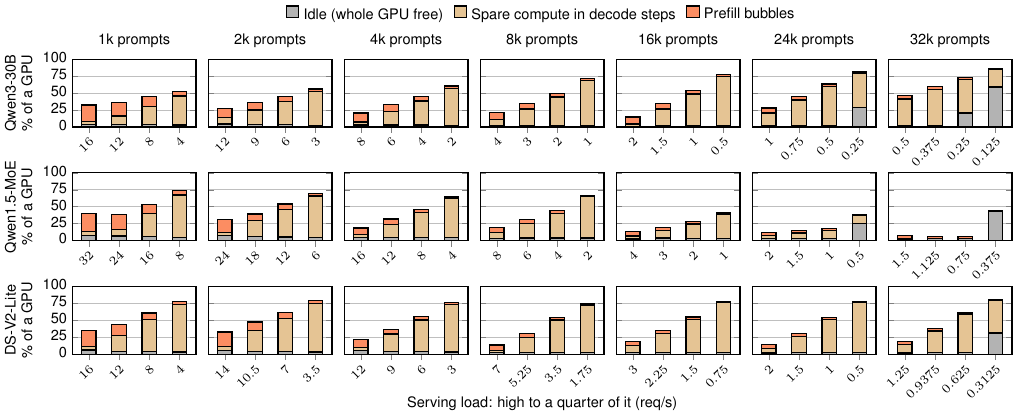}
  \caption{\textbf{Spare GPU capacity while serving alone}, as a share of
    one GPU: idle time (whole GPU free), spare compute inside decode
    steps that a co-running process can use without slowing them, and
    prefill bubbles. vLLM alone, \ac{TP}$=$\ac{EP}$=4$, 400 requests per
    run; x labels give the request rate, from the high load down to a
    quarter of it for each prompt length.}
  \label{fig:appx-capacity}
\end{figure*}

\S\ref{sec:eval-ablation} argues that bubbles matter more as serving load
consumes conventional headroom; here we measure that headroom directly.
Figure~\ref{fig:appx-capacity} decomposes the GPU capacity a colocated
job could use while vLLM serves alone, for every model, prompt length,
and load from a quarter of the high load up to the high load. Idle time
is wall time with no forward step running. Decode headroom is the
fraction of the GPU a co-running process can use inside decode steps
without measurably slowing them, measured with a throttled GEMM probe
under CUDA MPS. Prefill bubbles are the per-GPU waits before each layer's
all-reduce, inside which \sys{} opens windows for fine-tuning. For prompts up to 16k tokens, decode
headroom accounts for 35--74\% of a GPU at the lightest load but only
3--11\% at high load, while bubble capacity grows with prefill activity.

\begin{figure*}[t]
  \begin{minipage}[t]{\columnwidth}
    \vspace{0pt}\centering
  \inputplot{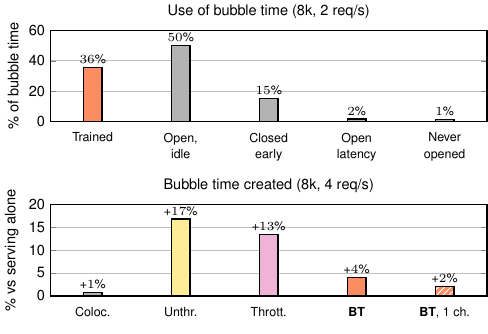}
  \caption{\textbf{Using expert-parallel bubbles.}
    Top: accounting of available non-straggler bubble time at
    8k tokens and 2 requests/s. Bottom: total bubble time created
    while different colocated execution policies run, normalized
    to serving alone, for \cfgColo (Coloc.), \cfgUnthr (Unthr.),
    \cfgThrottled (Thrott.), \sys{} (\sysshort), and \sys{} with one
    fine-tuning chain (\sysshort, 1 ch.); \S\ref{sec:eval-latency} defines
    the configurations. The accounting categories add up to 104\%: 46\,ms
    (3.8\% of bubble time) fall into two categories.}
  \label{fig:eval-bubble-use}
  \end{minipage}\hfill
  \begin{minipage}[t]{\columnwidth}
    \vspace{0pt}\centering
  \inputplot{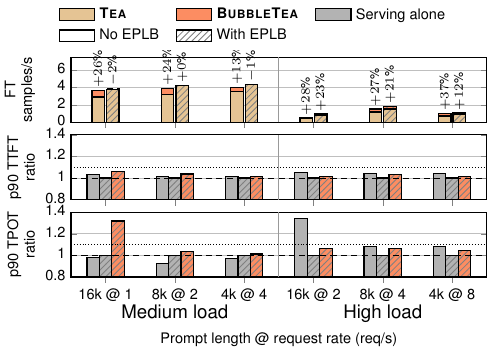}
  \caption{\textbf{\sys{} composes with expert load balancing, all
    workloads.}
    Top: fine-tuning throughput without and with \ac{EPLB}
    (hatched), split into \cfgOptColo{} and what bubble filling adds;
    labels give the change from filling, and a dashed outline marks a
    loss. Middle and bottom: p90 \ac{TTFT} and p90 \ac{TPOT} of
    serving alone and \sys{}, relative to serving alone with \ac{EPLB}
    (dashed line; dotted: 10\% increase).}
  \label{fig:appx-eplb}
  \end{minipage}
\end{figure*}

\begin{figure*}[t]
  \centering
  \inputplot{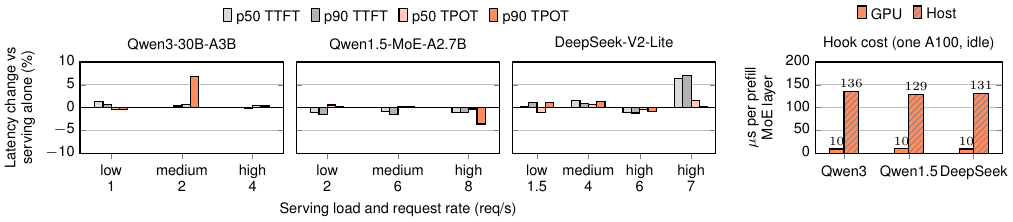}
  \caption{\textbf{Scheduling overhead with an idle fine-tuning worker.}
    Left three panels: p50 and p90 \ac{TTFT} and \ac{TPOT} with \sys{}'s
    hooks enabled but no training admitted, relative to serving alone,
    at 8k tokens and three loads per model (400 requests per run,
    pooled over runs). Right: per-layer cost of the prediction and
    admission hooks on one A100; solid bars are GPU time on the serving
    stream, hatched bars host time.}
  \label{fig:appx-sched-overhead}
\end{figure*}

\begin{figure*}[t]
  \centering
  \inputplot{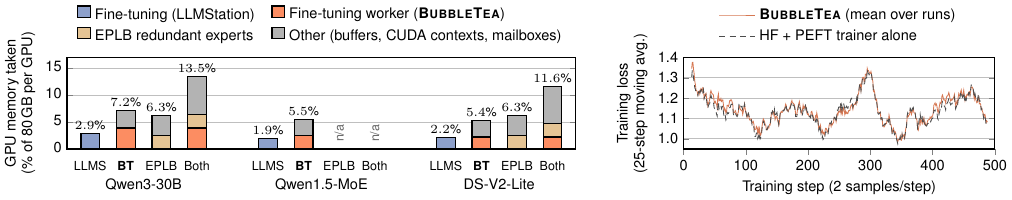}
  \caption{\textbf{Memory overhead and training correctness.}
    Left: GPU memory taken per GPU, as a share of 80\,GB, by LLMStation
    (LLMS), \sys{} (\sysshort), \ac{EPLB}'s redundant experts, and both together
    (Both).
    vLLM has no \ac{EPLB} for Qwen1.5-MoE. Right: 25-step moving-average
    training loss on Qwen3-30B-A3B for \sys{} and a standalone HF+PEFT
    trainer with the same data, LoRA configuration, and learning rate.}
  \label{fig:appx-overheads}

\end{figure*}

\begin{figure}[t]
  \centering
  \inputplot{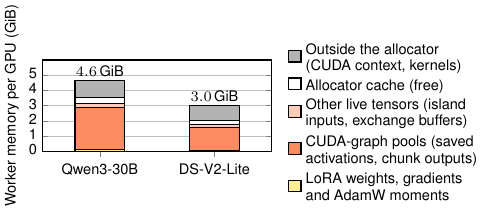}
  \caption{\textbf{Fine-tuning worker memory per GPU.} Breakdown of the resident
    \sys{} fine-tuning worker's footprint on Qwen3-30B-A3B and DeepSeek-V2-Lite.}
  \label{fig:appx-mem-breakdown}
\end{figure}

\subsection{Bubble utilization}
\label{sec:appx-bubble-use}

\S\ref{sec:eval-latency} notes that \sys{}'s protections leave some bubble
capacity unused; here we break down where that capacity goes. We define \emph{bubble utilization} as the fraction of measured
non-straggler bubble GPU-time during which fine-tuning computation
executes. Unlike the trace-driven experiment in Appendix~\ref{sec:appx-gap}, which
isolates what different planners could admit into recorded bubbles,
this measurement accounts for the complete \sys{} runtime: prediction,
admission, training dependencies, chunk readiness, and early close.

At 8k tokens and 2 requests/s, \sys{} executes useful fine-tuning work
during 36\% of the available bubble time. Approximately 50\% passes with
an admission window open but no fine-tuning chunk executing: the
fine-tuning worker is waiting on one of its own cross-GPU dependencies or has no ready
chunk. The remaining roughly 15\% is deliberately left unused because
\sys{} closes each window before its predicted end.

Open-but-idle time arises because \sys{} lets GPUs progress unevenly
(\S\ref{sec:design-compiler}) and does not try to equalize them.
Equalizing would rarely help: on Qwen3-30B-A3B at 16k tokens, the
most-loaded GPU changes between consecutive layers 70\% of the time,
and each of the four GPUs is the most-loaded in 14--39\% of layers,
so a GPU that falls behind in one layer often has a bubble to catch up in
the next, as well as during decode and idle time, which open on every
GPU. A GPU that runs ahead, however, can reach one of the training
step's cross-GPU exchanges before a lagging GPU has produced its
input, and its window then passes with nothing to run.

These losses have different causes. Open-but-idle time is a readiness
limitation: serving has exposed usable capacity, but training cannot
always supply work at the same fine granularity. Device-side
readiness and paired chains target this loss by keeping more work ready
(\S\ref{sec:design-chunk}), but they do not remove the dependencies
between fine-tuning chunks. Early-close time, in contrast, is capacity that
\sys{} gives up on purpose to avoid extending the serving critical path.

Colocated training could also create the capacity it later uses.
\cfgUnthr increases total bubble time by 17\% relative to serving
alone: training delays the most-loaded GPU, so its peers wait longer.
\sys{} increases it by only 4\% (2\% with one fine-tuning chain), because training
is admitted primarily on GPUs that would otherwise be waiting. The
windows also behave conservatively: 98\% close before the serving
dependency clears, with a median 57\,$\mu$s of remaining slack, and
the straggling GPU opens a window in only 4\% of layers.

\subsection{Composition with EPLB, all workloads}
\label{sec:appx-eplb}

\S\ref{sec:eval-eplb} shows that \sys{} composes with expert load
balancing at 8k prompts; here we report all six workloads.
Figure~\ref{fig:appx-eplb} extends Figure~\ref{fig:eval-eplb} to 4k, 8k,
and 16k prompts at medium and high load, with latency relative to
serving alone with \ac{EPLB}. The only workload outside 7\% of that
baseline is 16k prompts at 1 request/s, where p90 \ac{TPOT} reaches
1.32$\times$ (p90 \ac{TTFT} 1.06$\times$); there, bubble filling adds
no throughput ($-2$\%), so the residual bubbles are better left unused.
The tallest latency bar, 1.34$\times$ p90 \ac{TPOT} at 16k prompts and
2 requests/s, is serving alone \emph{without} \ac{EPLB}: it reflects
\ac{EPLB}'s own benefit to serving, not a cost of \sys{}, which stays
at 1.06$\times$ there.

\subsection{Overheads and correctness}
\label{sec:eval-overheads}

\S\ref{sec:evaluation} measures \sys{} with fine-tuning admitted; here we
isolate the cost of the runtime itself, independent of the fine-tuning
work it admits, and verify that its chunked execution preserves
training behavior.

\paragraph{Scheduling overhead.}
\sys{} places prediction and admission logic on the serving path, so
even a fine-tuning worker that never executes could in principle perturb
inference. To isolate this overhead, we run the complete runtime with
the fine-tuning worker resident but never admit a fine-tuning chunk. Each prefill
\ac{MoE} layer still predicts its bubble, opens an admission window,
arms early close, and closes the window.

Together, these hooks consume approximately 10\,$\mu$s of GPU time
per layer. Their host-side work takes approximately 130\,$\mu$s per
layer, but overlaps GPU execution; across all three models, measured
prefill-step duration changes by at most 0.5\%. Correspondingly, p50
and p90 \ac{TTFT} and \ac{TPOT} remain within 2.5\% of serving alone
in all but two configurations. On Qwen3-30B-A3B at medium load, p90
\ac{TPOT} rises by 6.8\% while p50 \ac{TPOT} and both \ac{TTFT}
percentiles stay within 1\%. The other exception is
DeepSeek-V2-Lite at its highest tested load, where the server is busy
approximately 95\% of the time and p50 and p90 \ac{TTFT} increase by
6.4\% and 7.0\%. Reducing the load from 7 to 6 requests/s removes this effect,
indicating that even sub-percent runtime overhead can amplify once the
server is already at the edge of queueing.
Figure~\ref{fig:appx-sched-overhead} reports the per-model results and
the per-layer hook cost.

\paragraph{Memory overhead.}
Keeping a separate fine-tuning worker resident reduces the memory available to the
serving KV cache by 4.3--5.8\,GiB per GPU, or 5--7\% of an 80\,GB A100,
compared with vLLM alone. The fine-tuning worker itself holds 3.0--4.6\,GiB
of this (Figure~\ref{fig:appx-mem-breakdown}); the rest is buffers,
CUDA contexts, and exchange mailboxes outside its allocator. The LoRA
parameters, gradients, and optimizer state take at most 0.15\,GiB;
most of the fine-tuning worker's memory is the activation pool backing the
compiled training graphs and its separate CUDA context. For comparison,
LLMStation reduces the KV cache by 1.5--2.4\,GiB per GPU because its
trainer executes inside the serving process and can share the CUDA
context and allocator. Figure~\ref{fig:appx-overheads} (left) compares
these reductions, including the additional cost when \ac{EPLB} is
enabled.

\paragraph{Training correctness and runtime reliability.}
\label{appendix:correctness}
Chunking a training step across short windows and two fine-tuning chains
must not change the optimization being performed. We therefore train
Qwen3-30B-A3B with \sys{} and with a standalone HF+PEFT trainer using
the same data, LoRA configuration, and learning rate. Across 476
steps, the 25-step moving-average losses differ by 1.2\% on average
and at most 4.5\%. We additionally observed 239,218 device-side
exchanges across 320 \sys{} runs with no mismatch, fallback, or
timeout (Figure~\ref{fig:appx-overheads}, right). Thus, the fine-grained execution required to exploit short
bubbles preserves the expected training behavior while
operating reliably across long-running experiments.

\section{Compiler passes}
\label{sec:appx-compiler}

Section~\ref{sec:design-compiler} summarizes how \sys{} derives
bubble-sized training work from the serving model; this appendix
lists the five passes.
\emph{Operation substitution} replaces each inference-only operation
with a differentiable one over the same weights: paged attention
becomes causal SDPA over the full sequence, MLA expands its latent keys
and values, fused \ac{MoE} becomes grouped \ac{GEMM}, and collectives
get explicit gradient rules. Scaling, positional encoding, routing, and
expert layout are kept, and LoRA deltas are inserted at the adapted
projections.
\emph{Graph capture} traces this forward with
\texttt{AOTAutograd} to obtain the backward graph, with the frozen base
weights as graph inputs so that the fine-tuning worker runs on the live serving
weights.
\emph{Exchange annotation} marks the partial gradients each GPU
produces for cross-GPU summation, with the ranks that take part.
\emph{Partitioning} separates forward from backward and compute from
communication, then uses a roofline cost model to cut compute into
chunks of at most 50\,$\mu$s on the target GPU. Cuts need not follow
layer or operator boundaries: the \ac{MoE} lowering splits the
token--expert pairs of one grouped \ac{GEMM} into several smaller ones.
\emph{Lowering} compiles compute chunks with \texttt{Inductor}, captures
sequences of gated chunks in CUDA graphs to avoid per-chunk launch
cost, and hands communication to \sys{}'s communicator.

\emph{Uneven progress is safe.}
Cross-GPU exchanges use one-sided mailbox messages tagged by exchange and round: a producer publishes without waiting, while consumers wait for the corresponding messages from their peers.
Because every GPU encounters exchanges in the same order and publishing never blocks, uneven progress cannot create a cyclic wait; a faster GPU can only wait until a slower peer reaches the same dependency.
Fine-tuning chains (\S\ref{sec:design-chunk}) synchronize when their gradients are combined for the optimizer step, which bounds how far GPUs drift apart and makes mailbox reuse safe.
Uneven progress may still leave some bubbles unused when a required peer result is not yet ready, but this affects utilization rather than correctness (Appendix~\ref{sec:appx-bubble-use}).

New models built from supported operations need no new code; the same
passes support Qwen1.5-MoE, Qwen3-MoE, DeepSeek-V2-Lite, and OLMoE.
A new operation needs a local lowering and a gradient rule, which can be
tested on its own (Appendix~\ref{appendix:correctness}).

\end{document}